\documentclass{aa}  
\usepackage[varg]{txfonts}
\usepackage{graphicx}
\usepackage{multirow}
\usepackage{hyperref}
\usepackage{natbib}
\usepackage{lscape}
\bibpunct{(}{)}{;}{a}{}{,} % to follow the A&A style

\usepackage{etoolbox}
\makeatletter
\patchcmd\@combinedblfloats{\box\@outputbox}{\unvbox\@outputbox}{}{%
 \errmessage{\noexpand\@combinedblfloats could not be patched}%
}%
\makeatother

\begin{document}

\title{The CARMENES search for exoplanets around M dwarfs}  
\subtitle{Activity indicators at visible and near-infrared wavelengths}
\titlerunning{Activity indicators}
\author{P.~Sch{\"o}fer\inst{1}
        \and
        S.~V.~Jeffers\inst{1}
        \and
        A.~Reiners\inst{1}
        \and
        D.~Shulyak\inst{2}
        \and
        B.~Fuhrmeister\inst{3}
        \and
        E.~N.~Johnson\inst{1}
        \and
        M.~Zechmeister\inst{1}
        \and
        I.~Ribas\inst{4,5}
        \and
        A.~Quirrenbach\inst{6}
        \and
        P.~J.~Amado\inst{7}
        \and
        J.~A.~Caballero\inst{8}
        \and
        G.~Anglada-Escud{\'e}\inst{7,9}
        \and
        F.~F.~Bauer\inst{7}
        \and
        V.~J.~S.~B{\'e}jar\inst{10,11}
        \and
        M.~Cort{\'e}s-Contreras\inst{8}
        \and
        S.~Dreizler\inst{1}
        \and
        E.~W.~Guenther\inst{12}
        \and
        A.~Kaminski\inst{6}
        \and
        M.~K{\"u}rster\inst{13}
        \and
        M.~Lafarga\inst{4,5}
        \and
        D.~Montes\inst{14}
        \and
        J.~C.~Morales\inst{4,5}
        \and
        S.~Pedraz\inst{15}
        \and
        L.~Tal-Or\inst{1,16}
        }
\institute{
        Institut f{\"u}r Astrophysik, Friedrich-Hund-Platz 1, D-37077 G{\"o}ttingen, Germany\\\email{schoefer@astro.physik.uni-goettingen.de}
        \and
        Max-Planck-Institut f{\"u}r Sonnensystemforschung, Justus-von-Liebig-Weg 3, D-37077 G{\"o}ttingen, Germany
        \and
        Hamburger Sternwarte, Gojenbergsweg 112, D-21029 Hamburg, Germany
        \and
        Institut de Ci{\`e}ncies de l'Espai (ICE, CSIC), Campus UAB, c/ de Can Magrans s/n, E-08193, Bellaterra, Barcelona, Spain
        \and
        Institut d'Estudis Espacials de Catalunya (IEEC), E-08034, Barcelona, Spain
        \and
        Landessternwarte, Zentrum f{\"u}r Astronomie der Universit{\"a}t Heidelberg, K{\"o}nigstuhl 12, D-69117, Heidelberg, Germany
        \and
        Instituto de Astrof\'{\i}sica de Andaluc\'{\i}a (IAA-CSIC), Glorieta de la Astronom\'{\i}a s/n, E-18008, Granada, Spain
        \and
        Centro de Astrobiolog\'{\i}a (CSIC-INTA), ESAC Campus, Camino Bajo del Castillo s/n, E-28692, Villanueva de la Ca\~{n}ada, Madrid, Spain
        \and
        School of Physics and Astronomy, Queen Mary, University of London, 327 Mile End Road, London, E1 4NS, UK
        \and
        Instituto de Astrof\'{\i}sica de Canarias, V\'{\i}a L{\'a}ctea s/n, 38205 La Laguna, Tenerife, Spain
        \and
        Departamento de Astrof\'{\i}sica, Universidad de La Laguna, E-38206 La Laguna, Tenerife, Spain
        \and
        Th{\"u}ringer Landessternwarte Tautenburg, Sternwarte 5, D-07778 Tautenburg, Germany
        \and
        Max-Planck-Institut f{\"u}r Astronomie, K{\"o}nigstuhl 17, D-69117, Heidelberg, Germany
        \and
        Departamento de Astrof\'{\i}sica y Ciencias de la Atm{\'o}sfera, Facultad de Ciencias F\'{\i}sicas, Universidad Complutense de Madrid, E-28040 Madrid, Spain
        \and
        Centro Astron{\'o}mico Hispano-Alem{\'a}n (CSIC-MPG), Observatorio Astron{\'o}mico de Calar Alto, Sierra de los Filabres, E-04550 G{\'e}rgal, Almer\'{\i}a, Spain
        \and
        School of Geosciences, Raymond and Beverly Sackler Faculty of Exact Sciences, Tel Aviv University, Tel Aviv, 6997801, Israel
        }
\date{Received 21 August 2018; accepted 18 January 2019}

% \abstract{}{}{}{}{} 
% 5 {} token are mandatory
\abstract
% context heading (optional), leave it empty if necessary 
{The Calar Alto high-Resolution search for M dwarfs with Exo-earths with Near-infrared and optical Echelle Spectrographs (CARMENES) survey is searching for Earth-like planets orbiting M dwarfs using the radial velocity method. Studying the stellar activity of the target stars is important to avoid false planet detections and to improve our understanding of the atmospheres of late-type stars.}
{In this work we present measurements of activity indicators at visible and near-infrared wavelengths for 331 M dwarfs observed with CARMENES. Our aim is to identify the activity indicators that are most sensitive and easiest to measure, and the correlations among these indicators. We also wish to characterise their variability.}
{Using a spectral subtraction technique, we measured pseudo-equivalent widths of the \ion{He}{i}~D$_3$, H$\alpha$, \ion{He}{i}~$\lambda$10833\,{\AA}, and Pa$\beta$ lines, the \ion{Na}{i}~D doublet, and the \ion{Ca}{ii} infrared triplet, which have a chromospheric component in active M dwarfs. In addition, we measured an index of the strength of two TiO and two VO bands, which are formed in the photosphere. We also searched for periodicities in these activity indicators for all sample stars using generalised Lomb-Scargle periodograms.}
{We find that the most slowly rotating stars of each spectral subtype have the strongest H$\alpha$ absorption. H$\alpha$ is correlated most strongly with \ion{He}{i}~D$_3$, whereas \ion{Na}{i}~D and the \ion{Ca}{ii} infrared triplet are also correlated with H$\alpha$. \ion{He}{i}~$\lambda$10833\,{\AA} and Pa$\beta$ show no clear correlations with the other indicators. The TiO bands show an activity effect that does not appear in the VO bands. We find that the relative variations of H$\alpha$ and \ion{He}{i}~D$_3$ are smaller for stars with higher activity levels, while this anti-correlation is weaker for \ion{Na}{i}~D and the \ion{Ca}{ii} infrared triplet, and is absent for \ion{He}{i}~$\lambda$10833\,{\AA} and Pa$\beta$. Periodic variation with the rotation period most commonly appears in the TiO bands, H$\alpha$, and in the \ion{Ca}{ii} infrared triplet.}
{}
\maketitle
%
%________________________________________________________________

\section{Introduction}
\label{section.introduction}
Because of their low masses and temperatures, M dwarfs have become interesting targets for radial velocity surveys, as Earth-like planets orbiting in their liquid-water habitable zones induce Doppler variations in the order of $1\,\mathrm{m\,s}^{-1}$. To reach this level of radial velocity precision, the Calar Alto high-Resolution search for M dwarfs with Exo-earths with Near-infrared and optical \'Echelle Spectrographs \citep[CARMENES,][]{2018SPIE10702E..0WQ} uses two spectrograph channels simultaneously that cover the visible wavelength range from $5200\,${\AA} to $9600\,${\AA} and the near-infrared wavelength range from $9600\,${\AA} to $17100\,${\AA}, where M dwarfs emit most of their flux.

The CARMENES sample consists of the brightest M dwarfs of each spectral subtype that are observable from Calar Alto \citep{2018A&A...612A..49R}. Despite the focus on finding planets, active stars were not removed from the sample, enabling us to study the stellar activity of a statistically significant sample.

One challenge for using the radial velocity method to find exoplanets around M dwarfs is that stellar activity can cause additional variations of the same order of magnitude as the signal induced by the planet \citep[e.g.][]{2014MNRAS.439.3094B}. It is therefore crucial to quantify the activity of the target stars in terms of strength and variability. Indicators derived from the cross-correlation function such as bisectors \citep{2001A&A...379..279Q} are closely connected to the effects of stellar activity on radial velocity measurements, whereas high-precision photometry traces starspots \citep[e.g.][]{2012MNRAS.419.3147A}. A third approach, which is directly related to the physical processes in the stellar atmosphere, is to examine spectral lines that are sensitive to activity. While the \ion{Ca}{ii} H\&K lines have been established as the standard activity indicator in Sun-like stars \citep[e.g.][]{1968ApJ...153..221W,1995ApJ...438..269B}, M dwarfs emit considerably less flux at these short wavelengths, which are therefore not covered by CARMENES. Late-type M dwarfs in particular are rarely included in \ion{Ca}{ii} H\&K surveys because they are faint at short wavelengths, and the H$\alpha$ line has instead been widely used to characterise the activity in samples of M dwarfs \citep[e.g.][]{1986ApJS...61..531S,2004AJ....128..426W,2018A&A...614A..76J}. Unlike \ion{Ca}{ii} H\&K, however, H$\alpha$ may give ambiguous results at low activity levels because it is expected to be an increasingly deep absorption line at first with increasing activity, then fill in, and finally go into emission \citep{1979ApJ...234..579C}.

In addition to H$\alpha$, other lines in the range of the visible channel of CARMENES that have been found to have chromospheric components sensitive to activity are the \ion{He}{i}~D$_3$ (hereafter He~D$_3$) line, the \ion{Na}{i}~D (hereafter Na~D) doublet, and the \ion{Ca}{ii} infrared triplet (hereafter Ca~IRT) \citep[e.g.][]{2002AJ....123.3356G,2009A&A...503..929H,2016ApJ...832..112R,2017A&A...605A.113M}. With the near-infrared channel, the \ion{He}{i}~$\lambda$10833\,{\AA} (hereafter He~10833) and Pa$\beta$ lines are accessible. They have also been investigated as chromospheric activity diagnostics in M dwarfs \citep[e.g.][]{1982ApJ...260..655Z,1998A&A...331L...5S,2008A&A...488..715S,2012ApJ...745...14S}. Another way to study the activity of M dwarfs is to examine titanium oxide and vanadium oxide absorption bands formed in the photosphere. Being sensitive to temperature, the strength of these bands can be affected by starspots. Moreover, the absorption bands of titanium oxide are sensitive to magnetic fields \citep{2002A&A...385..701B}, but also to metallicity \citep[e.g.][]{1982PASAu...4..417B,2007ApJ...669.1235L}, which is not related to activity.

In this paper we present an overview of pseudo-equivalent width measurements of the spectral lines that are sensitive to chromospheric activity at visible-light and near-infrared wavelengths using a spectral subtraction technique, and indices of photospheric TiO and VO bands for 331 stars of the CARMENES sample. We evaluate the sensitivity of these lines and investigate correlations among these activity indicators. Additionally, we explore the variations of these indicators for each star both in terms of the total variation and in view of periodicities.

The paper is structured as follows: we describe the observations in Sect.~\ref{section.observations}. In Sect.~\ref{section.indicators} we present our measurement methods. Section~\ref{section.coadded} gives an overview of the measured activity indicators and investigates how they are correlated, and Sect.~\ref{section.timeseries} explores how the activity indicators for each star vary in time. Finally, we summarise our results in Sect.~\ref{section.summary}.

%-------------------------------------------------------------------------------------------------------------------------------------------
\section{Observations}
\label{section.observations}
The CARMENES spectrograph is mounted on the $3.5\,$m telescope at the Calar Alto Observatory. A beam splitter in the telescope front-end separates the visible from the near-infrared light at a wavelength of $9600\,${\AA}. While both spectrograph channels operate independently, the same object can therefore be observed simultaneously with both spectrographs. The visible-light spectrograph provides a spectral resolution of $R=94600$ in the wavelength range from $5200\,${\AA} to $9600\,${\AA} with a mean sampling of $2.8\,$spectral bins per resolution element, and the near-infrared spectrograph covers the wavelength range from $9600\,${\AA} to $17100\,${\AA} with a spectral resolution of $R=80400$ and a mean sampling of $2.5\,$spectral bins per resolution element \citep{2018SPIE10702E..0WQ}.

The CARMENES sample is composed of the brightest stars for each spectral subtype from M0.0\,V to M9.0\,V that are observable from Calar Alto ($\delta > -23$\degr). Where available, we use the spectral types derived by \citet{2015A&A...577A.128A}, \citet{2013AJ....145..102L}, or \citet{1995AJ....110.1838R} from band indices in low-resolution spectra. For 30 stars not included in these catalogues, we use spectral types from various other sources \citep{1991ApJS...77..417K,2003AJ....125.1598L,2005A&A...442..211S,2003AJ....126.2048G,2006AJ....132..161G,2006AJ....132..866R,2007AJ....133.2258S,2009ApJ...704..975J,2009ApJ...699..649S,2010Ap.....53..123G,2014AJ....147...20N,2016Natur.534..658D,2017ApJ...834..187B}. While \citet{2017A&A...597A..47C} and \citet{2018A&A...614A..76J} identified binaries with a separation of less than $5\,$arcsec during the target preselection, \citet{2018A&A...619A..32B} discovered nine further double-lined spectroscopic binaries during the CARMENES survey. The separation of activity indicators in spectroscopic binaries is beyond the scope of this paper, and we therefore exclude these binaries. With this restriction, our sample consists of 331 stars, which are tabulated in Table~\ref{table:results}. It includes not only the 324 CARMENES survey stars presented by \citet{2018A&A...612A..49R}, but also 7 stars that were added to the survey later. As a consequence of selecting the brightest stars for each spectral subtype, our sample is focused on the solar neighbourhood and does not contain many stars from more distant and older populations. Of the 331 investigated stars, 98 stars were identified as members of the young disc population, 181 stars are in the thin disc, 13 stars in transition between thin and thick discs, and 23 stars in the thick disc \citep{2001MNRAS.328...45M,CC16}.

In this work, we analyse a total of 11000 visible and 10500 near-infrared spectra obtained between 3 January 2016 and 10 July 2018. The observations were not evenly distributed among the sample: while 103 stars were observed fewer than 10 times, 23 stars were observed more than 100 times. The spectra were reduced with the CARACAL pipeline using the flat-relative optimal extraction \citep{2014A&A...561A..59Z}. Additionally, a co-added spectrum of each star was computed from all individual spectra using SERVAL \citep{2018A&A...609A..12Z}. The co-added spectra provide an increased signal-to-noise ratio and were corrected for telluric lines, which appear in different positions in the individual spectra because of different barycentric velocities at different observation times. To explore the activity indicators across the sample, we derived an average value for each star from the co-added spectrum instead of averaging the indicators measured in individual spectra. This reduces the impact of outliers like strong flaring events, as data points outlying by more than $5\sigma$ are excluded from the co-adding. The results from individual spectra were used for investigating the variations in activity indicators for each star.

%-------------------------------------------------------------------------------------------------------------------------------------------
\section{Activity indicators}
\label{section.indicators}
In this section we describe the activity indicators and our measurement methods. As \ion{Ca}{ii} H\&K are not covered by CARMENES, we used the H$\alpha$ line as our main indicator of chromospheric activity. While \citet{2002AJ....123.3356G} reported a correlation between the emission strengths in H$\alpha$ and He~D$_3$, \citet{2009A&A...503..929H} noted that the He~D$_3$ line is in emission if there is emission in Na~D. Therefore, we included both the He~D$_3$ line and the Na~D doublet in our study. The Na~D doublet also has a strong photospheric absorption component that is sensitive to effective temperature, metallicity, and pressure, and it has therefore been used as an indicator for spectral type, metallicity, and luminosity class \citep{1923PASP...35..175L,1962ApJ...135..715S,2007AJ....134.2398C}. Towards the red end of the visible-light spectra, the Ca~IRT lines show chromospheric emission in active stars \citep{1997A&AS..124..359M,2017A&A...605A.113M}. Like the Na~D doublet, the Ca~IRT lines have strong photospheric absorption components that are sensitive to effective temperature, metallicity, and surface gravity \citep{1984ApJ...283..457J,1992A&A...254..258J}. We did not include the \ion{K}{i} $\lambda7700\,${\AA} and the \ion{Na}{i} $\lambda8200\,${\AA} doublets, which were investigated by \citet{2016ApJ...832..112R}, because they are located in spectral regions with strong telluric contamination. In the wavelength range covered by the near-infrared channel, \citet{2012ApJ...745...14S} observed the He~10833 and Pa$\beta$ lines in emission during M dwarf flares. The higher order lines of the Paschen series require higher excitation energies and therefore appear weaker, whereas the Pa$\alpha$ line is outside our spectral range.

Among the photospheric TiO absorption bands, the $\gamma$ band at $7050\,${\AA} and the $\epsilon$ band at $8430\,${\AA} are very sensitive to the effective temperature \citep{1993ApJ...402..643K}. Of the several photospheric VO absorption bands that have been used for spectral typing in late-M dwarfs and L dwarfs \citep{1999ApJ...519..802K,1999AJ....118.2466M} because of their sensitivity to cooler temperatures, we selected the bands at $7436\,${\AA} and $7942\,${\AA} because only few other spectral lines lie close to the band heads.

\subsection{Pseudo-equivalent widths}
\label{subsection.pEW}
As a measure of the chromospheric activity strength, we used the equivalent widths of the spectral lines selected above. Because M dwarf spectra do not show extensive continuum ranges, the equivalent width was defined with respect to a pseudo-continuum and is therefore called a pseudo-equivalent width ($pEW$).

To completely measure the emission caused by activity, we chose our integration windows $\lambda_0\pm\Delta\lambda$ to be sufficiently broad for the broadest emission lines seen in the analysed spectra. By using the integration windows listed in Table~\ref{table:lines}, we included most of the emission even in case of a strong flaring event. The disadvantage of broad integration windows is that they also include other nearby absorption lines. To minimise the effects of nearby lines and the photospheric absorption components of the Na~D and Ca~IRT lines, we used a spectral subtraction technique similar to what has been reported by \citet{1989ApJ...344..427Y} and \citet{1995A&A...294..165M}, and subtracted the spectrum of an inactive reference star.

\begin{table}
\caption{Integration windows and pseudo-continuum ranges for $pEW'$ measurements of the selected lines that are sensitive to chromospheric activity. All wavelengths are given in vacuum.}
\label{table:lines}
\centering
\begin{tabular}{l ccccc}
\hline\hline
Line &  $\lambda_0$ [\AA] & $\Delta\lambda$ [\AA] & $\lambda_\mathrm{l}$ [\AA] & $\lambda_\mathrm{r}$ [\AA] & $\Delta\lambda_\mathrm{PC}$ [\AA]\\    % table heading
\hline
He~D$_3$ & 5877.25 & 2.50 & 5872.75 & 5882.75 & 1.25\\
Na~D$_2$ & 5891.58 & 2.99 & \multirow{2}{*}{5882.75} & \multirow{2}{*}{5908.25} & \multirow{2}{*}{1.25}\\
Na~D$_1$ & 5897.56 & 2.99\\
H$\alpha$ & 6564.60 & 5.00 & 6552.50 & 6578.50 & 2.50\\
Ca~IRT-a & 8500.35 & 2.80 & 8494.00 & 8505.00 & 1.00\\
Ca~IRT-b & 8544.44 & 2.80 & 8538.50 & 8555.00 & 1.00\\
Ca~IRT-c & 8664.52 & 2.40 & 8658.00 & 8670.00 & 1.00\\
\hline
He~10833 & 10833.31 & 0.25 & 10821.0 & 10874.7 & 1.00\\
Pa$\beta$ & 12821.57 & 0.25 & 12817.0 & 12826.0 & 1.00\\
\hline
\end{tabular}
\end{table}

\begin{figure}
  \resizebox{\hsize}{!}{\includegraphics{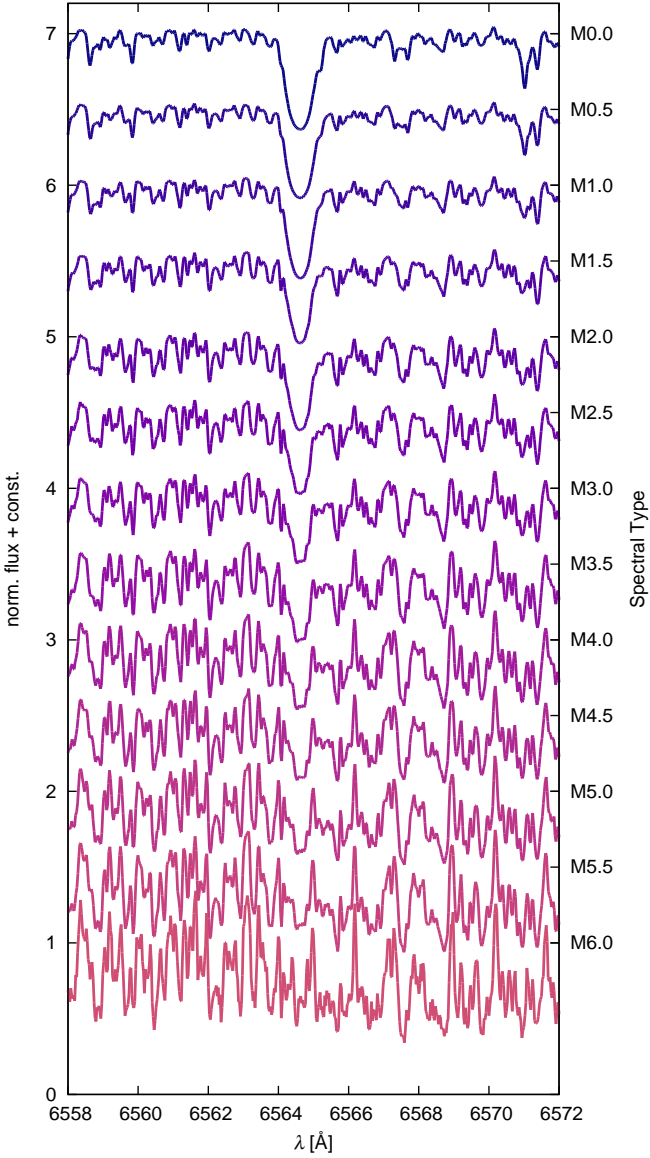}}
  \caption{H$\alpha$ region in the co-added spectra of the slow-rotating, H$\alpha$ inactive reference stars given in Table~\ref{table:refstars}.}
  \label{fig:Halpha_refspectra}
\end{figure}

We expect the strengths of photospheric absorption lines to depend mainly on the effective temperature, therefore we group our sample stars by spectral subtype. While metallicity and surface gravity also affect the photospheric absorption, especially in the cases of the Na~D and the Ca~IRT lines, we did not use a more detailed classification because this would lead to a large number of small groups that might not contain any inactive reference star, and the results for stars of different groups are not comparable without accounting for the different reference stars. Our reference star in each group was the star with the longest known rotation period $P_\mathrm{rot}$. We assumed that these are the least active stars of each spectral subtype because the relation between chromospheric activity and stellar rotation is well established \citep[e.g.][]{1998A&A...331..581D,2003ApJ...583..451M,2017ApJ...834...85N,2018A&A...614A..76J}. The disadvantage of this choice is that $P_\mathrm{rot}$ is known for only 154 of our 331 sample stars, so that our sample may include even slower rotators and hence less active stars. Our reference stars for spectral subtypes M0.5 and M2.0 have considerably shorter rotation periods than the other reference stars, so that it is likely that there are slower rotators with an undetected rotation period among the stars of these subtypes in our sample. On the other hand, this choice is independent of our measurements and does not rely on other measurements of the H$\alpha$ line. This is an advantage because according to \citet{1979ApJ...234..579C}, H$\alpha$ behaves ambiguously at low activity levels, as it first goes into deeper absorption before it is filled in and goes into emission with increasing activity strength. For spectral subtypes M5.5 and M6.0, however, all stars in our sample with a known rotation period have H$\alpha$ in emission, so that we instead selected the stars without H$\alpha$ emission as the reference stars for these subtypes. For stars of spectral subtype later than M6.0, we used the same reference star as for spectral type M6.0 because there are no very late-type stars without H$\alpha$ emission in our sample. All reference stars are listed in Table~\ref{table:refstars}. As shown in Fig.~\ref{fig:Halpha_refspectra}, all reference stars have H$\alpha$ in absorption, which indicates a minimum level of chromospheric heating that is necessary to excite hydrogen atoms. However, the reference stars do not show emission in the line wings, which we observe in stars with significantly less H$\alpha$ absorption. We therefore conclude that our reference stars are among the least active stars in our sample, although they have a minimum level of chromospheric activity.

\begin{table}
\caption{Reference stars for each spectral subtype. The same reference star is used for all spectral types later than M6.0\,V.}
\label{table:refstars}
\centering
\begin{tabular}{l c l c}
\hline\hline
Spectral type & $T_\mathrm{eff}$ [K] & Reference star & $P_\mathrm{rot}$ [d] \\    % table heading
\hline
M0.0\,V & 3850 & J14257+236W & 111 $\pm$ 12\\
M0.5\,V & 3800 & J18580+059 & 35.2 $\pm$ 0.3\\
M1.0\,V & 3680 & J18051--030 & 127.8 $\pm$ 3.2\tablefootmark{a}\\
M1.5\,V & 3600 & J16254+543 & 100 $\pm$ 5\\
M2.0\,V & 3550 & J06103+821 & 44.6 $\pm$ 1.0\\
M2.5\,V & 3450 & J17198+417 & 71.5 $\pm$ 2.6\\
M3.0\,V & 3400 & J15194--077 & 132.5 $\pm$ 6.3\tablefootmark{a}\\
M3.5\,V & 3250 & J17578+046 & 130\tablefootmark{b}\\
M4.0\,V & 3200 & J11477+008 & 163 $\pm$ 3\\
M4.5\,V & 3100 & J19216+208 & 133.5 $\pm$ 8.9\\
M5.0\,V & 3050 & J03133+047 & 126.2\tablefootmark{c}\\
M5.5\,V & 3000 & J00067--075 & \ldots\\
M6.0\,V & 2800 & J07403--174 & \ldots\\
M6.5\,V & 2700 & \ldots & \ldots\\
M7.0\,V & 2650 & \ldots & \ldots\\
M7.5\,V & 2610 & \ldots & \ldots\\
M8.0\,V & 2570 & \ldots & \ldots\\
M8.5\,V & 2510 & \ldots & \ldots\\
M9.0\,V & 2450 & \ldots & \ldots\\
M9.5\,V & 2400 & \ldots & \ldots\\
\hline
\end{tabular}
\tablefoot{Effective temperature estimates are adopted from \citet{2013ApJS..208....9P} and rotation periods from \citet{2018arXiv181003338D} if not indicated otherwise.\\
 \tablefoottext{a}{\citet{2015MNRAS.452.2745S}}
 \tablefoottext{b}{\citet{2007AcA....57..149K}}
 \tablefoottext{c}{\citet{2016ApJ...821...93N}}
 }
\end{table}

\begin{figure*}
  \resizebox{\hsize}{!}{\includegraphics{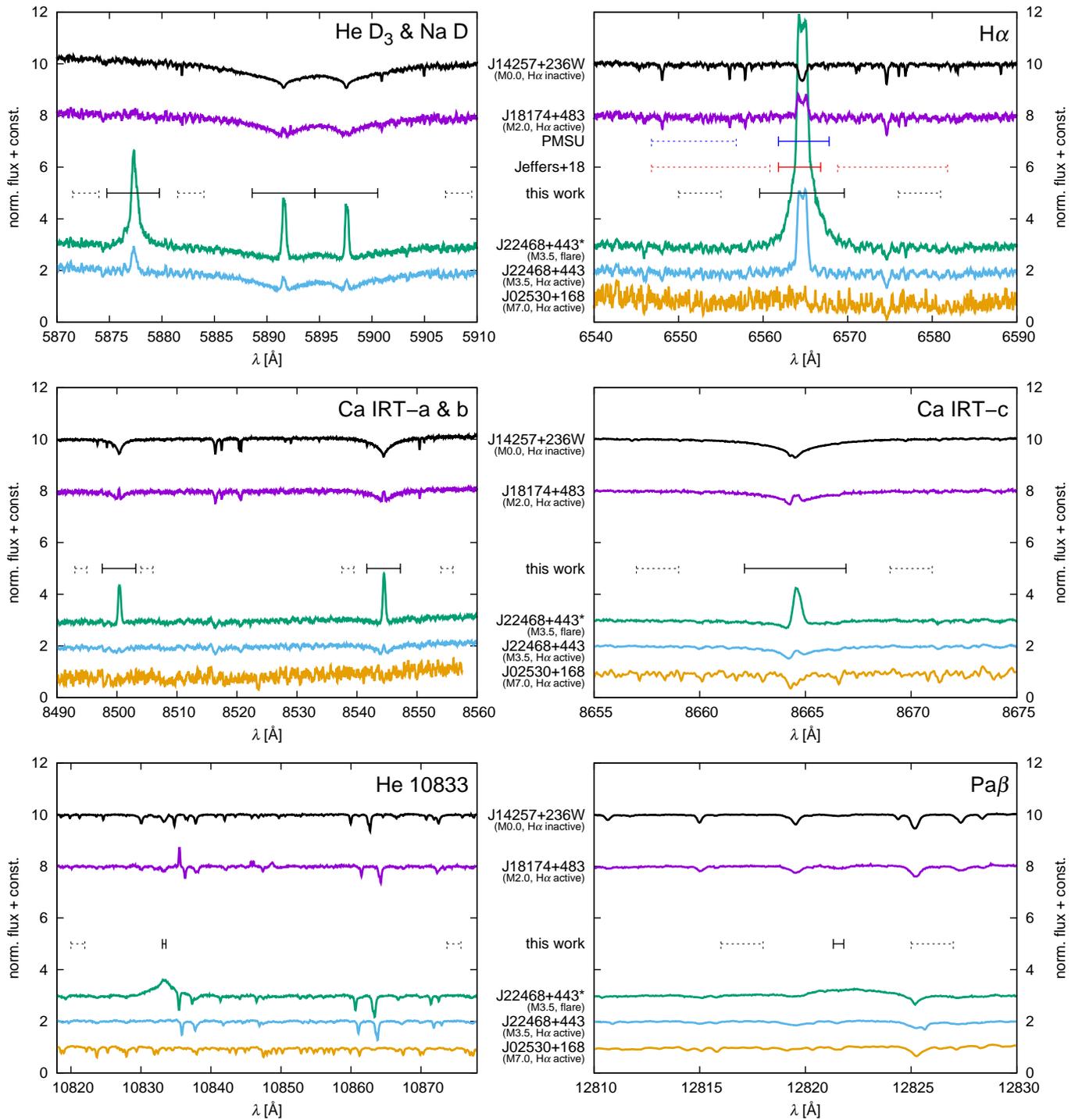}}
  \caption{Selected spectra in regions around the considered chromospheric lines: An inactive M0.0 star (J14257+236W, black), a moderately active M2.0 star (J18174+483, purple), a very active M3.5 star (J22468+443) in flaring (green) and quiescent state (blue), and a moderately active M7.0 star (J02530+168, orange). The region around the He~D$_3$ and Na~D lines (\textit{top left panel}) in the M7.0 spectrum is not shown because of a low signal-to-noise ratio. Bars mark the line window ranges (solid) and pseudo-continuum ranges (dashed) as given in Table~\ref{table:lines}. In the case of H$\alpha$, we also show the line window and pseudo-continuum ranges used by \citet{2018A&A...614A..76J} (red) and by \citet{1995AJ....110.1838R} and \citet{2015A&A...577A.128A} (blue).}
  \label{fig:specov_pEWs}
\end{figure*}

The subtraction of the co-added spectrum of the reference star from the investigated spectrum requires several preparation steps. First, we applied rotational broadening to the co-added spectrum of the reference star if the investigated star shows measurable rotational velocity values ($v\sin i$) above the detection limit of $2\,\mathrm{km\,s}^{-1}$ \citep{2018A&A...612A..49R}. Next, we transformed both spectra to a common wavelength grid by correcting for the Doppler shift between them. To determine the Doppler shift, we calculated the cross-correlation function of the spectral order that contains the considered spectral line and fit a Gaussian to the highest peak. After this, we normalised both spectra to their mean flux in the left and right pseudo-continuum ranges $\lambda_\mathrm{l/r}\pm\Delta\lambda_\mathrm{PC}$ located on either side of the considered line, which are given in Table~\ref{table:lines} and shown in Fig.~\ref{fig:specov_pEWs}. We then have the normalised, shifted spectrum $S(\lambda)$ and the normalised, broadened reference star spectrum $T(\lambda)$.

We define the pseudo-continuum $\mathrm{PC}(\lambda)$ as a linear interpolation between the pseudo-continuum ranges of $T(\lambda)$:
\begin{equation}
 \mathrm{PC}(\lambda) = \mathrm{PC}_\mathrm{l} + \frac{\mathrm{PC}_\mathrm{r} - \mathrm{PC}_\mathrm{l}}{\lambda_\mathrm{r} - \lambda_\mathrm{l}}\left(\lambda - \lambda_\mathrm{l}\right),
\end{equation}
where $\mathrm{PC}_\mathrm{l/r}$ is defined as the 90th percentile of $T(\lambda)$ in the pseudo-continuum ranges $\lambda_\mathrm{l/r}\pm\Delta\lambda_\mathrm{PC}$, which is robust to both the noise level and stellar absorption lines in the reference star spectrum.

For reasons of clarity, we denote the pseudo-equivalent width measured after the spectral subtraction as $pEW'$ and then calculated it as
\begin{equation}
 pEW' = -\int\limits_{\lambda_0-\Delta\lambda}^{\lambda_0+\Delta\lambda} \left(\frac{S(\lambda)}{\mathrm{PC}(\lambda)}-\frac{T(\lambda)}{\mathrm{PC}(\lambda)}\right)\,\mathrm{d}\lambda .
\end{equation}

We approximated the integral using the trapezoidal rule. To be consistent with the traditional definition of the equivalent width, positive values indicate absorption in the residual spectrum, while negative values indicate emission. We identified the inverse root mean square ($\mathit{rms}$) of the residual spectrum in the pseudo-continuum ranges as the signal-to-noise ratio and estimated the uncertainty of the pseudo-equivalent width using the formula derived by \citet{2006AN....327..862V}, which with our conventions and $\langle R\rangle$ the mean of the residual spectrum within the line window $\lambda_0\pm\Delta\lambda$ becomes
\begin{equation}
 \epsilon_{pEW'} = \sqrt{1+\frac{\mathrm{PC}(\lambda_0)}{\langle R\rangle+1}} \cdot \mathit{rms} \cdot \left(2\Delta\lambda - pEW'\right) .
\end{equation}
We note that $pEW'$ refers to the pseudo-equivalent width measured after the spectral subtraction as opposed to $pEW$ without spectral subtraction.

Despite the narrow integration windows for the He~10833 and Pa$\beta$ lines, telluric contamination depending on the Doppler shift of the stellar spectrum still remained. Therefore, we discarded $pEW'_{\mathrm{He}~10833}$ measurements where the observed absolute radial velocity including the barycentric motion of Earth was between $14\,\mathrm{km\,s}^{-1}$ and $125\,\mathrm{km\,s}^{-1}$ and $pEW'_{\mathrm{Pa}\beta}$ measurements where it was between $10\,\mathrm{km\,s}^{-1}$ and $43\,\mathrm{km\,s}^{-1}$.

\subsection{Photospheric band indices}
\label{subsection.indices}
\begin{figure*}
  \resizebox{\hsize}{!}{\includegraphics{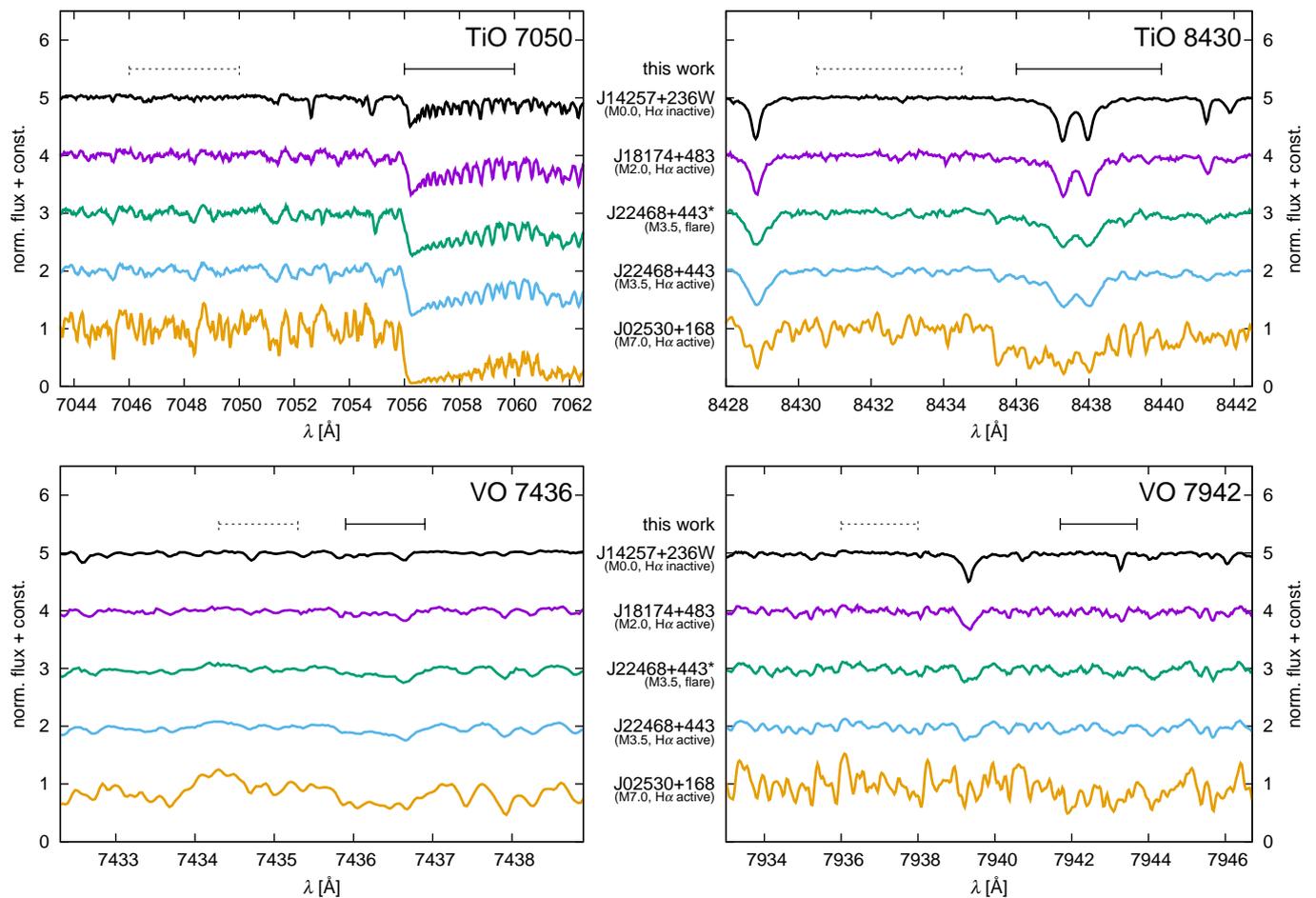}}
  \caption{Same spectra as in Fig.~\ref{fig:specov_pEWs} in regions around the considered photospheric bands. Bars mark the numerator ranges (solid) and denominator ranges (dashed) as given in Table~\ref{table:bands}.}
  \label{fig:specov_indices}
\end{figure*}

In case of the photospheric TiO and VO absorption bands, the pseudo-equivalent width is not a good measure to quantify the strength because they may extend over multiple spectral orders and are blended with other spectral features. Instead, we measured indices derived from the mean fluxes in two small ranges on either side of the band head:
\begin{equation}
 index = \frac{\langle S\rangle_\mathrm{num.}}{\langle S\rangle_\mathrm{den.}} ,
\end{equation}
with the numerator and denominator ranges given in Table~\ref{table:bands} and shown in Fig.~\ref{fig:specov_indices}. Similar indices have been used previously to measure the strength of photospheric absorption bands \citep[e.g.][]{1995AJ....110.1838R,1999ApJ...519..802K}. However, they were derived for low-resolution spectra using greater numerator and denominator ranges. Because these ranges also include other spectral features, the resulting indices are a less accurate measure of the absorption band strength.

Following \citet{2018A&A...609A..12Z}, we estimated the error of the mean fluxes from the data error $\epsilon_i$ of each bin in the considered spectral range as
\begin{equation}
 \epsilon_{\langle S\rangle} = \frac{1}{N}\sqrt{\sum\epsilon_i^2}
\end{equation}
and calculated the uncertainty of the index by propagating these errors:
\begin{equation}
 \epsilon_{index} = index \sqrt{\left(\frac{\epsilon_\mathrm{num.}}{\langle S\rangle_\mathrm{num.}}\right)^2+\left(\frac{\epsilon_\mathrm{den.}}{\langle S\rangle_\mathrm{den.}}\right)^2}
.\end{equation}
Using these formulations, we calculated the activity indicators for individual observations and the co-added spectra for all 331 stars in our sample.

\begin{table}
\caption{Numerator and denominator ranges for photospheric absorption band indices. All wavelengths are given in vacuum.}
\label{table:bands}
\centering
\begin{tabular}{l cc}
\hline\hline
Band &  Numerator [\AA] & Denominator [\AA]\\    % table heading
\hline
TiO 7050 & 7056.0:7060.0 & 7046.0:7050.0\\
VO 7436 & 7435.9:7436.9 & 7434.3:7435.3\\
VO 7942 & 7941.7:7943.7 & 7936.0:7938.0\\
TiO 8430 & 8436.0:8440.0 & 8430.5:8434.5\\
\hline
\end{tabular}
\end{table}

%-------------------------------------------------------------------------------------------------------------------------------------------
\section{Analysis of results from co-added spectra}
\label{section.coadded}
This section analyses the activity indicators derived from the co-added spectra that are listed in Table~\ref{table:results}. We show the spectral type dependence of the activity indicators and compare our results to those of previous M dwarf activity studies. In addition, we investigate the correlations among the chromospheric indicators and the effects of activity on the photospheric bands.

\subsection{Spectral-type dependence of activity indicators}
\label{subsection.sptdependence}
As an overview of our measured quantities, we show the spectral type dependence of the $pEW'$ and indices derived from the co-added spectra for each of the 331 sample stars. This serves to demonstrate the effect of the spectral subtraction technique in the $pEW'$ calculation and to motivate the investigation of correlations among the different indicators.

\subsubsection{Chromospheric lines}
\label{subsubsection.sptdpnd.chromospheric}
\begin{figure*}
  \resizebox{\hsize}{!}{\includegraphics{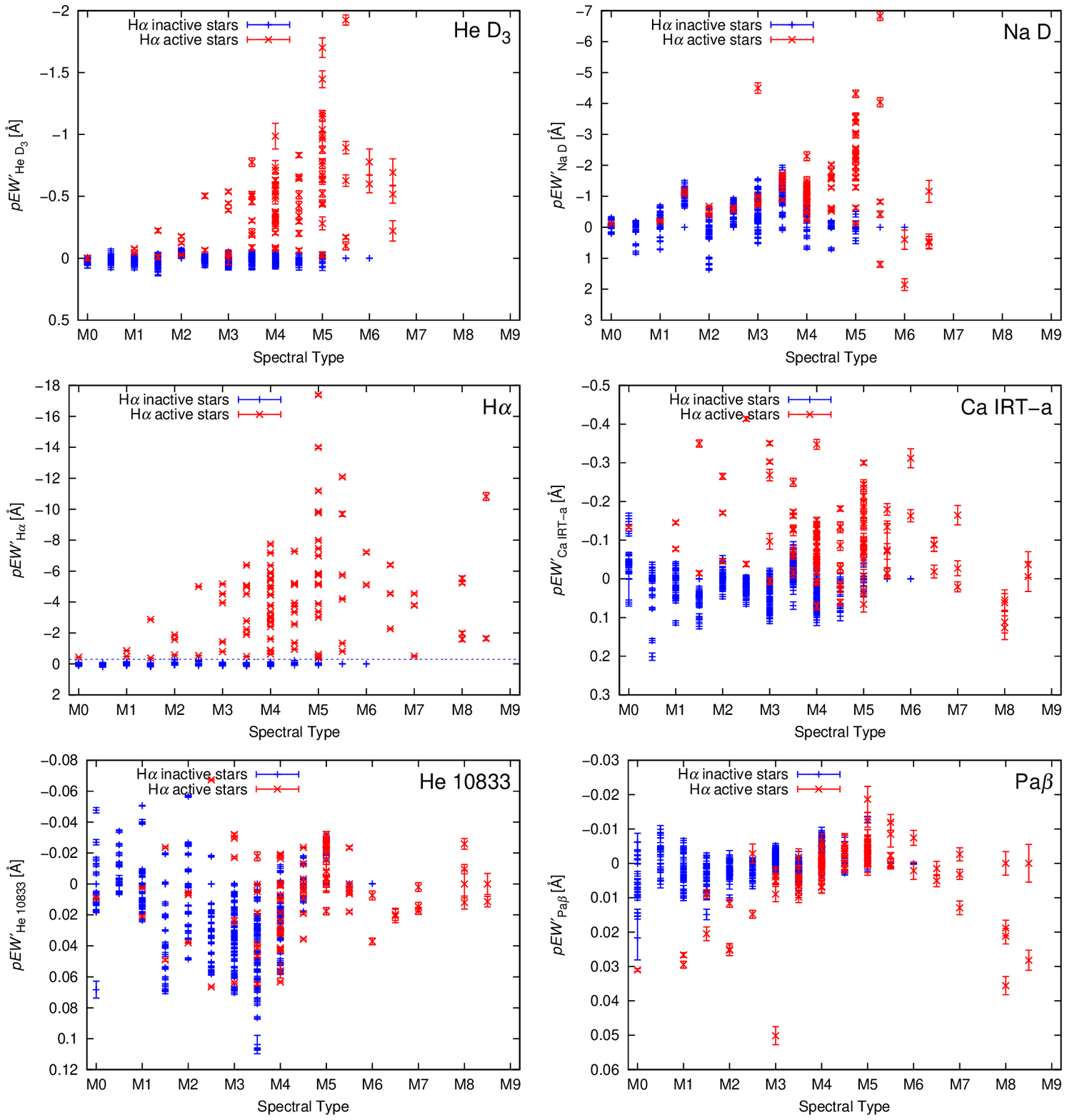}}
  \caption{$pEW'$ of chromospheric lines as a function of spectral type. $pEW'_\mathrm{Na~D}$ is the sum of the Na~D$_1$ and Na~D$_2$ lines. The dashed line in the left middle panel marks our activity threshold of $pEW'_{\mathrm{H}\alpha} = -0.3\,${\AA}. Ca~IRT-b and -c are similar to Ca~IRT-a and therefore not shown. The reference star for each spectral subtype up to M6.0 has $pEW' = 0\,${\AA} for all lines by definition.}
  \label{fig:pEW_vs_SpT}
\end{figure*}

We plot the $pEW'$ of the chromospheric lines as a function of spectral type in Fig.~\ref{fig:pEW_vs_SpT}. We note that by definition, the reference stars used for the spectral subtraction technique have $pEW' = 0\,${\AA} for all lines. Stars with $pEW'_{\mathrm{H}\alpha} < -0.3\,${\AA} are marked as H$\alpha$ active. This threshold is considerably different from previous M dwarf activity studies, which defined activity thresholds of $-0.5\,${\AA} \citep{2018A&A...614A..76J}, $-0.75\,${\AA} \citep{2011AJ....141...97W}, or $-1\,${\AA} \citep{2017ApJ...834...85N}, for instance,  for $pEW_{\mathrm{H}\alpha}$ without spectral subtraction, that is,emission with respect to the pseudo-continuum rather than excess emission with respect to a reference star with H$\alpha$ absorption. However, we find that relatively few stars have $pEW'_{\mathrm{H}\alpha}\approx -0.3\,${\AA}, and therefore small variations or a slightly different activity threshold do not significantly change the number of H$\alpha$ active stars. The threshold of $pEW'_{\mathrm{H}\alpha} < -0.3\,${\AA} is further justified in Sect.~\ref{subsection.variability} below.

For He~D$_3$, we find that all H$\alpha$ inactive stars have values comparable to the reference star of their spectral subtype, whereas most H$\alpha$ active stars have a negative $pEW'_\mathrm{He~D_3}$, that is, excess emission in He~D$_3$ compared to the respective reference star. The maximum observed He~D$_3$ excess emission increases with the spectral subtypes up to M5.5 and declines at later subtypes. The He~D$_3$ line is not included in the co-added spectra of stars with spectral subtype M7.0 or later because the flux in that part of the spectrum is too low. Hence, $pEW'_\mathrm{He~D_3}$ cannot be measured for these very late-type stars.

In the case of Na~D, we show the summed results for the Na~D$_1$ and Na~D$_2$ lines. We find that the H$\alpha$ inactive stars of different spectral subtypes are not closely grouped around the reference stars and cover different ranges of $pEW'_\mathrm{Na~D}$ values. On the one hand, no trend is visible for these groups, so that the spectral type dependence of the photospheric Na~D component is removed by the spectral subtraction technique. On the other hand, a new bias is introduced by the reference star selection because of the metallicity and surface gravity dependence, as is clearly visible for spectral subtype M1.5. The selected reference star would be an outlier with excess absorption in Na~D if a different star were chosen as the reference. The H$\alpha$ active stars in general have Na~D excess emission. Their $pEW'_\mathrm{Na~D}$ values are similar to the values of the H$\alpha$ inactive stars of the same spectral subtype for spectral subtypes up to M4.0, whereas the Na~D excess emission is stronger at later spectral subtypes. As with He~D$_3$, there are no Na~D doublet measurements for stars with spectral subtype M7.0 or later.

For H$\alpha$, we find that all H$\alpha$ inactive stars have very similar $pEW'_{\mathrm{H}\alpha}$ values. In particular, there are no stars with considerably stronger H$\alpha$ absorption than the reference star of the respective spectral subtype. This implies that the least active stars have a certain minimum level of chromospheric activity because the strongest H$\alpha$ absorption according to the predictions by \citet{1979ApJ...234..579C} as described in Sect.~\ref{subsection.pEW} would translate into $pEW'_{\mathrm{H}\alpha}\approx 0.6\,${\AA} if the reference star had a completely inactive chromosphere. We observe that the $pEW'_{\mathrm{H}\alpha}$ of the H$\alpha$ active stars is typically more negative at later spectral subtypes, with a maximum at spectral subtype M5.0. In part, this is caused by the lower pseudo-continuum level at later spectral subtypes, as we show in Sect.~\ref{subsection.comparison}.

The results for the Ca~IRT lines are similar to each other, so that we only show Ca~IRT\nobreakdash-a. We find that the $pEW'_\mathrm{Ca~IRT-a}$ values of the H$\alpha$ inactive stars are distributed in a similar way as the Na~D results. As with Na~D, the spectral type dependence of the photospheric component of the Ca~IRT\nobreakdash-a line is removed by the spectral subtraction technique. However, the reference star selection may introduce a bias caused by differences in metallicity and surface gravity. We measured a Ca~IRT\nobreakdash-a excess emission for most H$\alpha$ active stars. In contrast to the He~D$_3$, Na~D, and H$\alpha$ lines, the excess emission is typically strongest at spectral subtypes M4.0 and earlier. This is an effect of the temperature-dependent flux ratio between the pseudo-continua at the wavelengths of the Ca~IRT lines and at the shorter wavelengths of the He~D$_3$, Na~D, and H$\alpha$ lines.

For He~10833, we find that the H$\alpha$ inactive stars are distributed in different ways depending on the spectral subtype. At spectral subtypes M0.0 to M1.0 and M4.5, about the same number of H$\alpha$ inactive stars have excess absorption as have excess emission with respect to the reference star. However, at spectral subtypes M1.5 to M4.0, excess absorption occurs more often, whereas at spectral subtype M5.0, excess emission occurs more often. The H$\alpha$ active stars are not clearly separated from the H$\alpha$ inactive stars.

In the case of Pa$\beta$, we find no clear trends among the H$\alpha$ inactive stars. The H$\alpha$ active stars of spectral subtypes earlier than M4.0 have excess absorption, whereas the H$\alpha$ active stars of later spectral subtypes are not clearly separated from the H$\alpha$ inactive stars and some have excess emission.

Comparing the results for the different lines, we find that He~D$_3$ and H$\alpha$ show a very similar behaviour. In Na~D and Ca~IRT\nobreakdash-a, the photospheric line components introduce an additional spread to the results. The He~10833 and Pa$\beta$ lines completely differ from the other lines and are more challenging to measure because we are limited to narrow integration windows by telluric lines and the co-added near-infrared spectra are not as reliably cleaned from telluric lines as the co-added visible-light spectra.

\subsubsection{Photospheric bands}
\label{subsubsection.sptdpnd.photospheric}
\begin{figure*}
  \resizebox{\hsize}{!}{\includegraphics{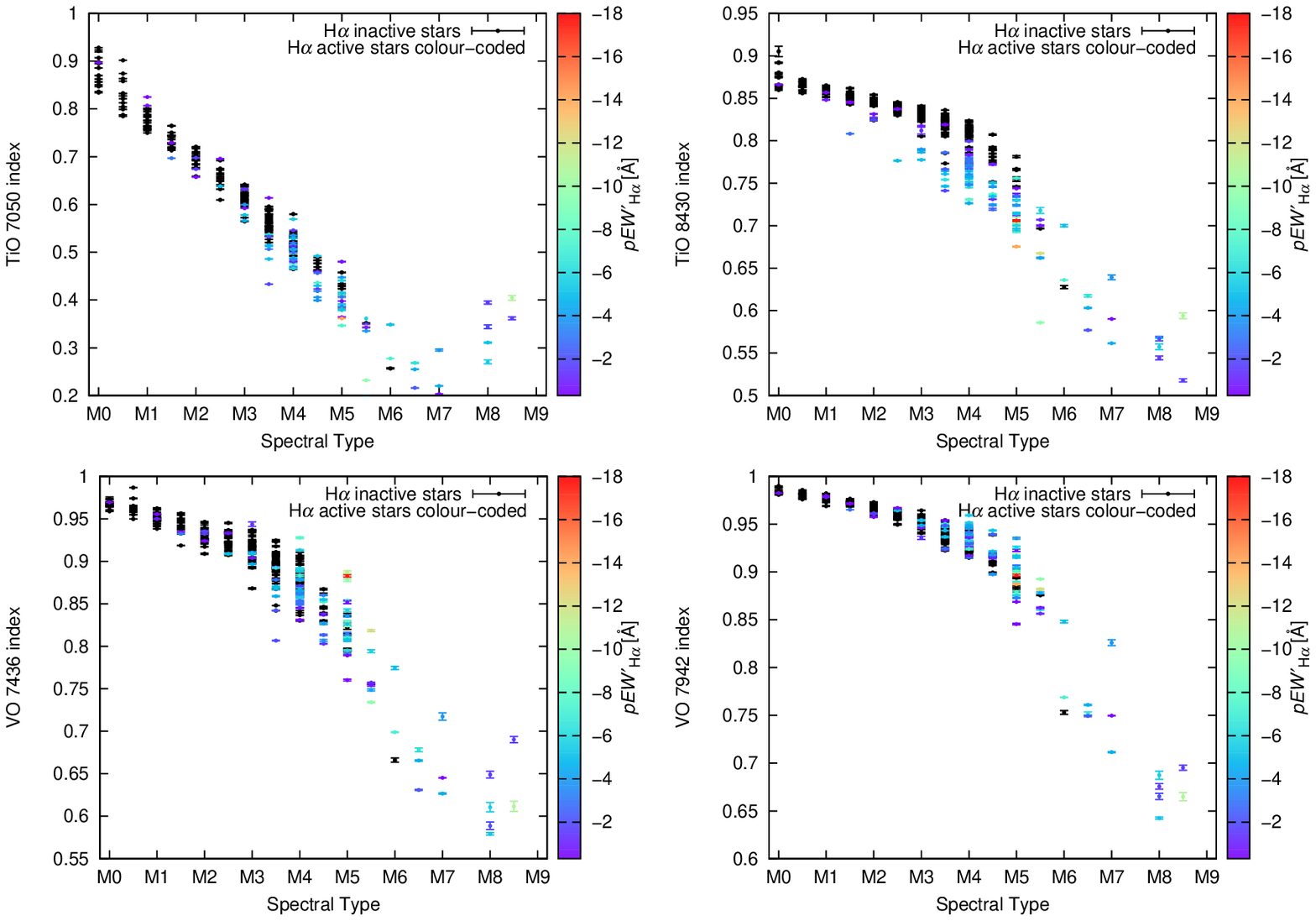}}
  \caption{Photospheric line indices TiO~7050 (\textit{top left}), TiO~8430 (\textit{top right}), VO~7436 (\textit{bottom left}), and VO~7942 (\textit{bottom right}) as a function of spectral type with H$\alpha$ active stars coloured corresponding to their $pEW'_{\mathrm{H}\alpha}$.}
  \label{fig:index_vs_SpT}
\end{figure*}

In Fig.~\ref{fig:index_vs_SpT} we show our photospheric TiO and VO band indices as a function of the spectral subtype. As these indices are calculated without the spectral subtraction, the spectral type dependence of each index is clearly visible. The H$\alpha$ active stars are colour-coded according to their $pEW'_{\mathrm{H}\alpha}$ value.

For TiO~7050, there is an almost linear decrease of the index with the spectral subtype up to M6.5, indicating that the absorption band is stronger at lower effective temperatures. At spectral subtypes later than M6.5, the index increases again. Our TiO~7050 index is very similar to the TiO 2 index of \citet{1995AJ....110.1838R}, for which \citet{2015A&A...577A.128A} reproduced the same behaviour. The TiO~8430 index also decreases with the spectral type, but while there is only a small decrease from spectral subtype M0.0 to M3.0, it decreases more rapidly at the later spectral types. We also note that the H$\alpha$ active stars generally have lower TiO index values than H$\alpha$ inactive stars of the same spectral subtype. This is discussed in more detail in Sect.~\ref{subsection.photosphericcorr} below.

The VO~7436 index behaves similarly to the TiO~8430 index with a small decrease for spectral subtypes up to M3.0 and a stronger decrease for the later spectral types. \citet{2002AJ....123.3409H} reported the same behaviour for their VO~7434 index, which measures the strength of the same absorption band using different spectral ranges. VO~7942 also shows a similar dependence, but the more rapid decrease starts at a later spectral subtype around M4.0.

\subsection{Comparison to previous M dwarf activity studies}
\label{subsection.comparison}
\begin{figure}[h!]
  \resizebox{\hsize}{!}{\includegraphics{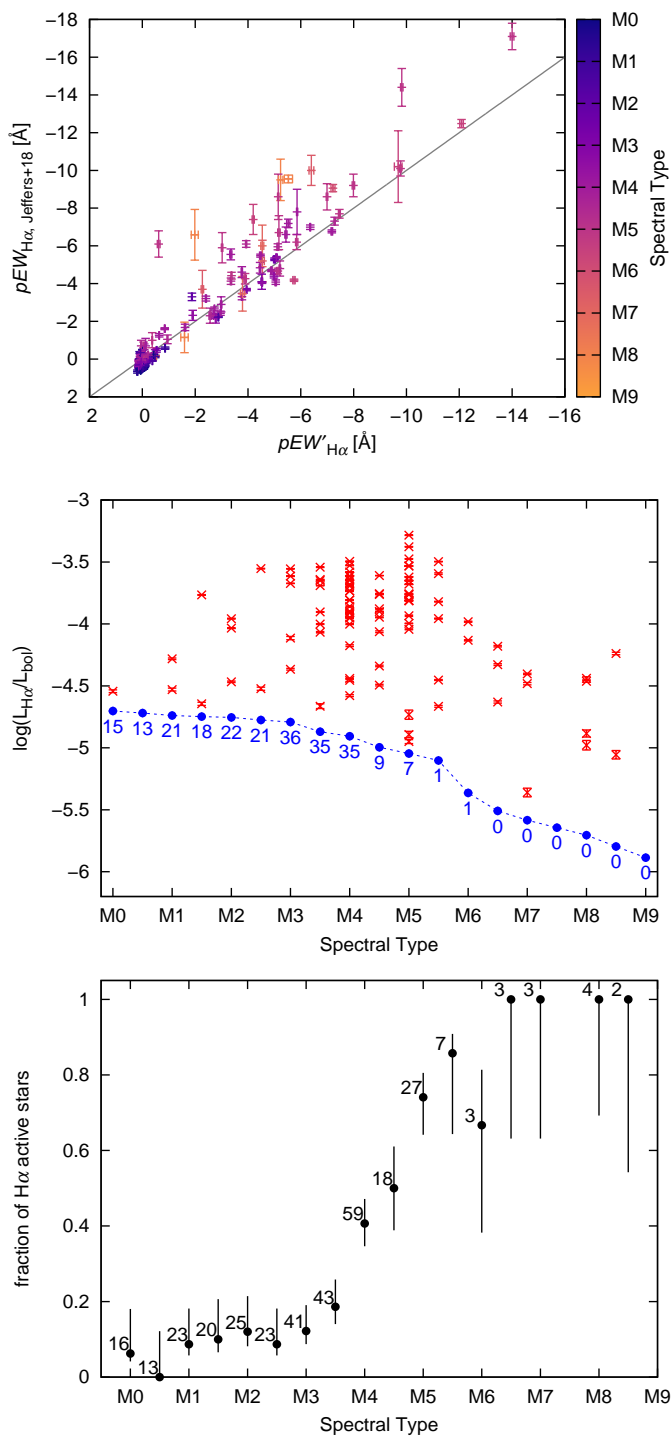}}
  \caption{\textit{Top}: Comparison of our $pEW'_{\mathrm{H}\alpha}$ values with the $pEW_{\mathrm{H}\alpha}$ values from \citet{2018A&A...614A..76J}. Colours denote spectral types. \textit{Middle}: Normalised H$\alpha$ luminosity of active stars as a function of spectral type (red crosses). The blue circles indicate our activity threshold of $pEW'_{\mathrm{H}\alpha} = -0.3\,${\AA,} and the numbers reflect the number of stars below this threshold. \textit{Bottom}: Fraction of H$\alpha$ active stars per spectral subtype for the 331 stars of our sample with error bars from binomial statistics. Numbers reflect the total number of sample stars in the corresponding bin.}
  \label{fig:Halpha_comp_ov}
\end{figure}

Previous investigations of the chromospheric activity indicators other than H$\alpha$ focused on earlier type stars \citep[e.g.][]{2017A&A...605A.113M} or individual M dwarfs \citep{2016ApJ...832..112R}. Hence, a meaningful comparison is only possible for H$\alpha$.  \citet{2018A&A...614A..76J}  presented a catalogue of $pEW_{\mathrm{H}\alpha}$ for approximately 2200 M dwarfs, including all our 331 sample stars, using different instruments. In addition to our own measurements, we collected measurements from \citet{1995AJ....110.1838R}, \citet{2015A&A...577A.128A}, \citet{2014MNRAS.443.2561G}, \citet{2015AJ....149..106D}, \citet{2013AJ....145..102L}, \citet{2006AJ....132..866R}, \citet{2015ApJ...812....3W}, \citet{2017ApJ...834...85N}, \citet{2010ApJ...710..924R}, \citet{2014AJ....147...85R}, and \citet{2010ApJ...716.1522S}. In contrast to this work, we used narrower integration windows and different pseudo-continuum ranges, which are shown in the top right panel of Fig.~\ref{fig:specov_pEWs} for comparison, and we did not use the spectral subtraction technique. Since all our sample stars are covered by \citet{2018A&A...614A..76J}, we compare our $pEW'_{\mathrm{H}\alpha}$ measurements only with the $pEW_{\mathrm{H}\alpha}$ values from that catalogue in the top panel of Fig.~\ref{fig:Halpha_comp_ov}.

The group of stars around $pEW'_{\mathrm{H}\alpha} = 0\,${\AA}, which has H$\alpha$ in absorption, shows the effect of our spectral subtraction technique compared to $pEW_{\mathrm{H}\alpha}$ measured without spectral subtraction. Although H$\alpha$ absorption is stronger in earlier spectral type stars, our results are close to $pEW'_{\mathrm{H}\alpha} = 0\,${\AA} for all spectral types. As we measured the H$\alpha$ emission above the absorption line of a reference star rather than above the pseudo-continuum, our values might be expected to be more negative than the literature values. Different line windows and definitions of the pseudo-continuum cause an opposite effect, however, so that the literature values are in general more negative than our results. In addition, the literature values were derived from observations at different times and using different instruments, so that true variations in the emission strength and instrumental effects may cause more deviations. We note, however, that for every star the order of magnitude of our $pEW'_{\mathrm{H}\alpha}$ and the literature $pEW_{\mathrm{H}\alpha}$ is the same.

As noted in Sect.~\ref{subsubsection.sptdpnd.chromospheric}, the $pEW'$ is a measure of the line height or depth in relation to the pseudo-continuum and therefore not directly comparable across different spectral types because the absolute level of the pseudo-continuum changes with the effective temperature. A more comparable measure is the luminosity observed in an emission line, which we calculated for the H$\alpha$ line normalised to the bolometric luminosity $L_\mathrm{bol}$ by multiplying $pEW'_{\mathrm{H}\alpha}$ with the ratio between the pseudo-continuum flux and the bolometric flux $\chi$ \citep{2004PASP..116.1105W} when $pEW'_{\mathrm{H}\alpha}$ was below our activity threshold of $-0.3\,${\AA},
\begin{equation}
 \frac{L_{\mathrm{H}\alpha}}{L_\mathrm{bol}} = -\frac{pEW'_{\mathrm{H}\alpha}}{1\,\textrm{\AA}} \cdot \chi(T_\mathrm{eff}) .
\end{equation}
To compute the flux ratio $\chi(T_\mathrm{eff})$, we used a quintic function derived from PHOENIX model spectra by \citet{2008ApJ...684.1390R}. Modelling $\chi(T_\mathrm{eff})$ for the other chromospheric emission lines is beyond the scope of this paper, therefore we only performed this calculation for H$\alpha$. While \citet{2018A&A...615A...6P} derived the effective temperature $T_\mathrm{eff}$ from the CARMENES co-added spectra used in this work, a single $T_\mathrm{eff}$ measurement of our reference star might not be representative for the corresponding spectral subtype. Hence, we used the $T_\mathrm{eff}$ estimates from \citet{2013ApJS..208....9P} as given in Table~\ref{table:refstars}, which are based on several different measurements for well-established reference stars of each spectral subtype.

As shown in the middle panel of Fig.~\ref{fig:Halpha_comp_ov}, while the normalised H$\alpha$ luminosities of most active stars with a spectral type of M2.5 or earlier are barely above our detection limit, significantly more stars have $\log(L_{\mathrm{H}\alpha}/L_\mathrm{bol})$ between $-3.5$ and $-4$ at spectral types between M3.5 and M5.5 with a maximum at M5.0. For the later types, the typical normalised H$\alpha$ luminosity decreases more strongly than the typical $pEW'_{\mathrm{H}\alpha}$. This is in agreement with the result from \citet{2018A&A...614A..76J} for a similar sample. In the bottom panel of Fig.~\ref{fig:Halpha_comp_ov}, we show the fraction of stars per spectral subtype that we identified as H$\alpha$ active with 67\% credibility ranges from binomial statistics. While the fraction is below 20\% at spectral types up to M3.5, it notably increases between M4.0 and M4.5, which marks the transition into the fully convective regime \citep{2011ASPC..448..505S}. With the exception of our reference stars, all stars with a spectral type later than M5.0 are H$\alpha$ active. Within the error bars, this agrees with the previous results of \citet{2012AJ....143...93R}, \citet{2015ApJ...812....3W}, and \citet{2018A&A...614A..76J} (among others) for similar samples of nearby M dwarfs even though different measurement methods and activity thresholds were used. While inactive stars of spectral types later than M6.0 do exist \citep[e.g.][]{2008AJ....135..785W,2013AJ....146...50P}, they belong to older populations at larger distances and are therefore too faint to be included in our sample. Overall, we find that 96 (29\%) of our 331 sample stars are H$\alpha$ active.

\subsection{Correlations among chromospheric lines}
\label{subsection.chromosphericcorr}
\begin{figure*}
  \resizebox{\hsize}{!}{\includegraphics{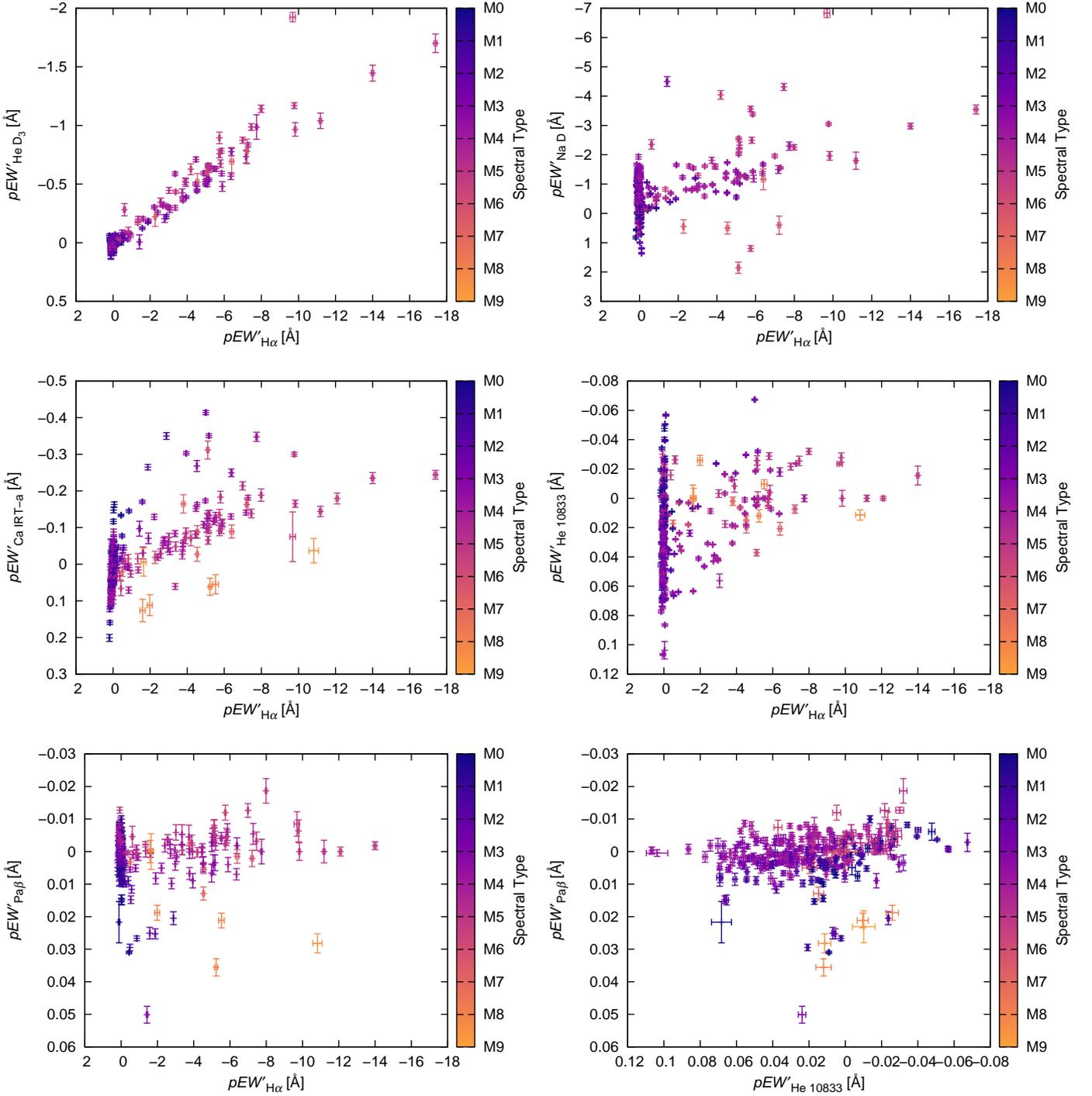}}
  \caption{Scatter plots of $pEW'$ values of H$\alpha$ and He~D$_3$ (\textit{top left}), H$\alpha$ and Na~D (\textit{top right}), H$\alpha$ and Ca~IRT-a (\textit{centre left}), H$\alpha$ and He~10833 (\textit{centre right}), H$\alpha$ and Pa$\beta$ (\textit{bottom left}), and He~10833 and Pa$\beta$ (\textit{bottom right}). Colours denote spectral types.}
  \label{fig:pEW_vs_pEW}
\end{figure*}

The spectral type dependences of our activity indicators discussed in Sect.~\ref{subsection.sptdependence} suggest that  different chromospheric activity indicators are correlated. In this section, we test whether the chromospheric He~D$_3$, Na~D, Ca~IRT\nobreakdash-a, He~10833, and Pa$\beta$ lines are correlated with H$\alpha$. Using the $pEW'$ values derived from the co-added spectra for the 331 sample stars, we computed the Spearman correlation coefficients $r_S$ between H$\alpha$ and each of the other lines as well as between He~10833 and Pa$\beta$. The corresponding scatter plots are shown in Fig.~\ref{fig:pEW_vs_pEW}.

We find that He~D$_3$ is in excess emission for all stars where $pEW'_{\mathrm{H}\alpha} < -0.6\,${\AA}. Because the He~D$_3$ excess emission increases linearly with further increasing H$\alpha$ excess emission, we find a strong correlation with $r_S = 0.70$. The correlation coefficient increases to $r_S = 0.96$ when we consider only the H$\alpha$ active stars. A linear correlation between these two lines was previously reported by \citet{2002AJ....123.3356G}. Although the flux ratio of the pseudo-continua depends on the effective temperature, we do not find a significant dependence of the slope on spectral type.

In the case of Na~D, the H$\alpha$ inactive stars can have excess absorption or emission with respect to the reference stars, as was described in Sect.~\ref{subsubsection.sptdpnd.chromospheric}. This results in an elongated cloud of data points at $pEW'_{\mathrm{H}\alpha} > -0.3\,${\AA}. We find only a weak correlation with $r_S = 0.20$ for all sample stars. For the H$\alpha$ active stars, however, the excess emission in H$\alpha$ and Na~D is correlated with $r_S = 0.60$. Again, we do not find a spectral type dependence.

For Ca~IRT\nobreakdash-a, we also find an elongated cloud of data points formed by the H$\alpha$ inactive stars and stronger Ca~IRT\nobreakdash-a excess emission for increasing H$\alpha$ excess emission in general. However, while the correlation coefficient is $r_S = 0.73$ for all sample stars, the correlation is weaker among the H$\alpha$ active stars with $r_S = 0.61$. This is a consequence of the spectral type dependence of the slope, which we find to be steeper for earlier spectral types. To determine whether the steeper slopes are an effect of the different pseudo-continua, we estimated the continuum radiation at different spectral types by approximating the stars as blackbodies with temperature $T_\mathrm{eff}$ from Table~\ref{table:refstars}. Planck's law yields that at the wavelength of H$\alpha$, an M0.0 star ($T_\mathrm{eff} = 3850\,$K) has a continuum radiation of $330\,\mathrm{W}\,\mathrm{m}^{-2}\,\mathrm{sr}^{-1}\,${\AA}$^{-1}$, whereas an M4.0 star ($T_\mathrm{eff} = 3200\,$K) has a continuum radiation of $104\,\mathrm{W}\,\mathrm{m}^{-2}\,\mathrm{sr}^{-1}\,${\AA}$^{-1}$. Therefore, the same excess energy $\Delta_{\mathrm{H}\alpha}$ results in a $pEW'_{\mathrm{H}\alpha}$ that is 3.17 times as large for an M4.0 star as for an M0.0 star. In contrast, the continuum radiation at the wavelength of Ca~IRT\nobreakdash-a is $335\,\mathrm{W}\,\mathrm{m}^{-2}\,\mathrm{sr}^{-1}\,${\AA}$^{-1}$ for an M0.0 star and $136\,\mathrm{W}\,\mathrm{m}^{-2}\,\mathrm{sr}^{-1}\,${\AA}$^{-1}$ for an M4.0 star. Therefore, the same excess energy $\Delta_{\mathrm{Ca~IRT-a}}$ results in a $pEW'_{\mathrm{Ca~IRT-a}}$ that is 2.46 times as high for an M4.0 star as for an M0.0 star. Consequently, a constant energy ratio $\Delta_{\mathrm{H}\alpha}/\Delta_{\mathrm{Ca~IRT-a}}$ results in a $pEW'$ ratio that for an M4.0 star is only 0.78 times as high as for an M0.0 star. This explains the fainter slope for later spectral types that we find in the plot.

He~10833 can show excess emission or excess absorption for the H$\alpha$ inactive stars. We do not see a clear correlation with $pEW'_{\mathrm{H}\alpha}$ for the H$\alpha$ active stars. The correlation coefficient is only $r_S = 0.27$ for all sample stars, and $r_S = 0.41$ for the H$\alpha$ active stars. However, with increasing H$\alpha$ excess emission, fewer stars have He~10833 excess absorption. There is no clear spectral type dependence.

In the case of Pa$\beta$, we find excess emission or excess absorption for the H$\alpha$ inactive stars as well, no clear correlation with $pEW'_{\mathrm{H}\alpha}$ for the H$\alpha$ active stars, and no spectral type dependence of the correlation. Pa$\beta$ excess absorption is more common in stars with $pEW'_{\mathrm{H}\alpha} > -6\,${\AA}, whereas more stars with stronger H$\alpha$ excess emission have Pa$\beta$ excess emission. The correlation coefficient is $r_S = -0.16$ for all sample stars and $r_S = 0.42$ for the H$\alpha$ active stars. There is no strong correlation between Pa$\beta$ and He~10833 either ($r_S = 0.26$ for all sample stars, $r_S = 0.44$ for the H$\alpha$ active stars).

The correlations are in agreement with our observations in Sect.~\ref{subsubsection.sptdpnd.chromospheric}. H$\alpha$ is correlated best with He~D$_3$, while the correlations with Na~D and Ca~IRT\nobreakdash-a are weaker because the photospheric component of these lines is not completely removed by the spectral subtraction technique. The near-infrared He~10833 and Pa$\beta$ lines show no strong monotonic correlation with H$\alpha$ or with each other. We do not see any clear nonmonotonic correlation, but cannot rule out that there are outliers caused by telluric lines in the integration windows of these two lines.

\subsection{Activity effects on photospheric bands}
\label{subsection.photosphericcorr}
As noted in Sect.~\ref{subsection.sptdependence}, our photospheric TiO and VO absorption band indices show not only the expected spectral type dependence, but also a different behaviour with increasing H$\alpha$ excess emission. Figure~\ref{fig:index_vs_SpT} shows the indices as a function of the spectral subtype, colour-coded according to $pEW'_{\mathrm{H}\alpha}$. The H$\alpha$ active stars in general have lower TiO~7050 index values than the H$\alpha$ inactive stars of the same spectral subtype. This effect is stronger at later spectral subtypes. \citet{1996AJ....112.2799H} and \citet{2015A&A...577A.128A} reported the same behaviour for their TiO 2 index, which is comparable to our TiO~7050 index. The separation between the H$\alpha$ active and H$\alpha$ inactive stars appears stronger in TiO~8430, whereas no clear separation is visible in the two VO band indices.

The spectral types were derived from empirical functions of spectral indices similar to our photospheric band indices \citep[e.g.][]{1995AJ....110.1838R,2013AJ....145..102L,2015A&A...577A.128A}. This means that the spectral types may be biased depending on the set of spectral indices used for the spectral type determination. In addition, spectral types are no exact quantities and the $T_\mathrm{eff}$ ranges covered by different spectral subtypes can overlap. The direct determination of $T_\mathrm{eff}$, however, also uses photospheric bands like the TiO $\gamma$ band at $7050\,${\AA} \citep{2018A&A...615A...6P} and may in consequence also be biased by other effects. Therefore we study the photospheric band indices only as a function of the spectral type. Because younger stars tend to be more active \citep{1972ApJ...171..565S,2008AJ....135..785W}, it is possible that the H$\alpha$ active stars in our sample are younger and more metal rich than the H$\alpha$ inactive stars, and that the differences in the photospheric band indices are caused by different metallicities. While most H$\alpha$ active stars in our sample belong to the young disc or thin disc population, there are also H$\alpha$ inactive stars in both populations, and we do not see any significant difference in the distribution of photospheric band indices within each population. The metallicities measured by \citet{2015A&A...577A.128A} and \citet{2018A&A...615A...6P} do not show any correlation with the photospheric band indices for stars of the same spectral type either. Because population membership is only a very rough age estimate, and the metallicity measurements again are based on photospheric bands and therefore may be affected by temperature differences and other effects, we cannot rule out that the differences in the photospheric band indices of H$\alpha$ active and H$\alpha$ inactive stars are a metallicity effect. We do not have evidence for a metallicity effect either, however, therefore we discuss other differences between active and inactive stars that may affect the photospheric bands in the following.

One difference between active and inactive stars are active regions such as starspots. The temperature dependence of TiO bands has been used to measure starspot areas and temperatures \citep{1995ApJ...452..879N,1998ApJ...507..919O}. However, a lower apparent temperature of the stellar surface affects the whole spectrum and thus results in a later spectral type. Moreover, the VO band indices show a spectral-type and hence temperature dependence comparable to the TiO~8430 index for the H$\alpha$ inactive stars and should thus show the same behaviour if there was a temperature change. Starspots therefore cannot explain our observations.

\begin{figure}
  \resizebox{\hsize}{!}{\includegraphics{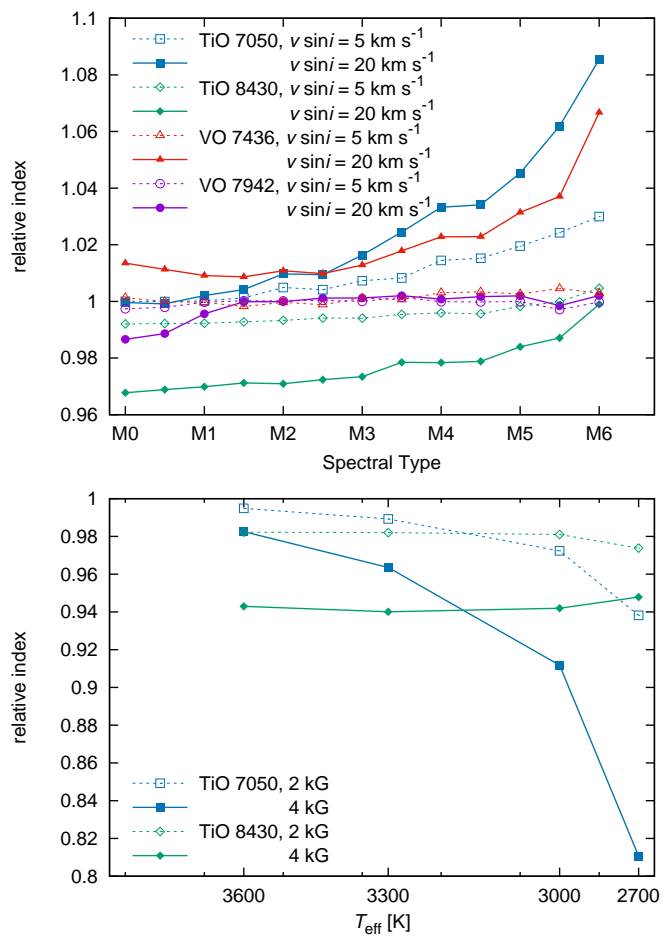}}
  \caption{\textit{Top}: Effect of different rotational velocities ($v\sin i$) on the photospheric band indices as measured by the ratio of the index of the co-added spectra of the reference stars with an artificial rotational broadening to the index without rotational broadening as a function of the spectral type. \textit{Bottom}: Effect of magnetic fields on the TiO band indices as measured by the ratio of the index of model spectra with homogeneous magnetic fields of different strengths to the index of model spectra without a magnetic field for different $T_\mathrm{eff}$. The $T_\mathrm{eff}$ scale is nonlinear to be consistent with the spectral type scale in the top panel.}
  \label{fig:index_vsini_mf}
\end{figure}

Another difference between active and inactive stars is the faster rotation of more active stars \citep[e.g.][]{1998A&A...331..581D,2003ApJ...583..451M,2017ApJ...834...85N,2018A&A...614A..76J}. To investigate the effect of stellar rotation on our photospheric band indices, we applied an artificial rotational broadening to the co-added spectra of our reference stars and measured the indices in the broadened spectra. The upper panel of Fig.~\ref{fig:index_vsini_mf} shows the indices derived in this way for $v\sin i=5\,\mathrm{km\,s}^{-1}$ and $v\sin i=20\,\mathrm{km\,s}^{-1}$ divided by the indices without the rotational broadening as a function of the spectral type. While the TiO~7050 index increases with higher $v\sin i$ and this increase is larger at later spectral subtypes, the TiO~8430 index decreases with higher $v\sin i$ and the effect is smaller at later spectral subtypes. In contrast, a $v\sin i$ of $5\,\mathrm{km\,s}^{-1}$ has no significant effect on either VO band index and a high $v\sin i$ of $20\,\mathrm{km\,s}^{-1}$ increases the VO~7436 index more strongly at later spectral subtypes, whereas the VO~7942 is only affected at early spectral subtypes where no fast rotators exist in our sample. Faster rotation of more active stars can therefore explain the behaviour of the TiO~8430 and the two VO band indices, but it cannot be the only relevant effect because it results in higher TiO~7050 index values for active stars than for inactive stars, while we observe lower values.

A third difference between active and inactive stars may be stronger magnetic fields in the active stars, which lead to Zeeman broadening of the individual lines in the band. We investigated the effect of magnetic fields on the two TiO band indices using model spectra calculated for different $T_\mathrm{eff}$ without magnetic fields and with homogeneous magnetic fields with different strengths of $2\,$kG and $4\,$kG. The spectra were computed with the \texttt{MagneSyn} magnetic spectrum synthesis code described in \citet{2017NatAs...1E.184S}. The transition parameters of TiO lines were extracted from the latest release of the VALD database \citep{2015PhyS...90e4005R}. As the rotation-spin coupling constants for the VO bands are unknown, we cannot study how magnetic fields affect our VO band indices. The lower panel of Fig.~\ref{fig:index_vsini_mf} shows the ratio of the TiO indices of the model spectra with a magnetic field divided by the TiO indices of the model spectrum without a magnetic field for the same $T_\mathrm{eff}$. We find that the TiO~7050 index decreases with stronger magnetic fields and the decrease is larger at lower $T_\mathrm{eff}$. The TiO~8430 index also decreases with stronger magnetic fields, but this effect does not depend on $T_\mathrm{eff}$. The effect of Zeeman broadening on the indices is consistent with our observations.

While the TiO~8430 index is decreased by both rotational and Zeeman broadening, the TiO~7050 index is decreased by Zeeman broadening but increased by rotational broadening. This shows that a combination of both effects can explain that the observed separation between the H$\alpha$ active and the H$\alpha$ inactive stars of the same spectral type is stronger in the TiO~8430 index. It may also explain why the H$\alpha$ active stars were not assigned later spectral subtypes, as the spectral types were derived from low-resolution spectra where the effects of rotational and Zeeman broadening may be diminished.

%-------------------------------------------------------------------------------------------------------------------------------------------
\section{Time series of activity indicators}
\label{section.timeseries}
We now consider individual observations rather than the co-added spectra. Using the time series of the activity indicators, we investigate their variation quantitatively and examine if the variation is periodic.

\subsection{Variability of activity indicators}
\label{subsection.variability}
We defined the absolute variation of each indicator as the difference between the 20th and 80th percentiles, which we calculated using the quantile estimator developed by \citet{1982Biom...69..635H}. This definition is more suitable for our data than the variance or standard deviation because occasional strong flaring events can lead to a non-normal distribution of the measured values for each star. By using the quantile estimator rather than direct percentiles, we also obtain the errors of the percentiles without assuming a specific distribution. We propagated these errors to calculate the error of the absolute variation and tabulate the results in Table~\ref{table:results}.

\begin{figure}
  \resizebox{\hsize}{!}{\includegraphics{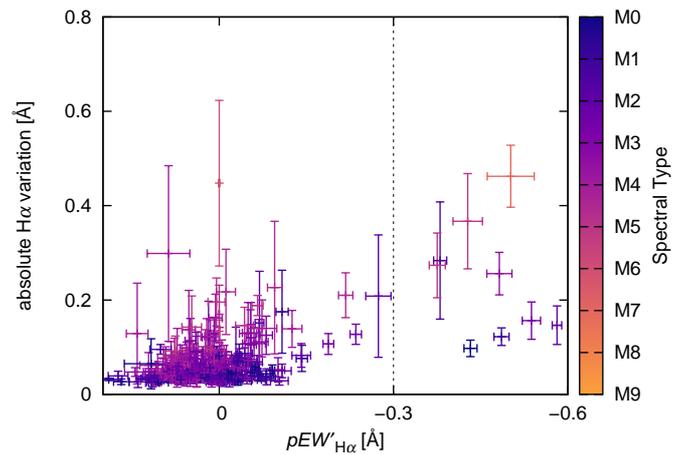}}
  \caption{Absolute H$\alpha$ variation as defined by the difference between the 20th and 80th percentiles as a function of $pEW'_{\mathrm{H}\alpha}$ for $pEW'_{\mathrm{H}\alpha} \approx 0\,${\AA}. Colours denote spectral types.}
  \label{fig:Halpha_var}
\end{figure}

The absolute H$\alpha$ variation is shown as a function of $pEW'_{\mathrm{H}\alpha}$ for $pEW'_{\mathrm{H}\alpha} \approx 0\,${\AA} in Fig.~\ref{fig:Halpha_var}. While the H$\alpha$ inactive stars show a mean variation of $0.07\,${\AA} in $pEW'_{\mathrm{H}\alpha}$, the variations of the H$\alpha$ active stars range between $0.10\,${\AA} and $8.4\,${\AA}. In particular, the stars with H$\alpha$ excess emission slightly stronger than our activity threshold of $pEW'_{\mathrm{H}\alpha} = -0.3\,${\AA} show larger H$\alpha$ variations than most H$\alpha$ inactive stars, strengthening the choice of this threshold. It is also visible that the variations among the H$\alpha$ inactive stars are larger for later spectral types. This is expected because a lower signal-to-noise ratio leads to larger statistical fluctuations. In contrast, true variability of the stellar activity is the dominant source of H$\alpha$ variations in H$\alpha$ active stars because their variations are typically an order of magnitude larger than the measurement errors.

\begin{figure*}
  \resizebox{\hsize}{!}{\includegraphics{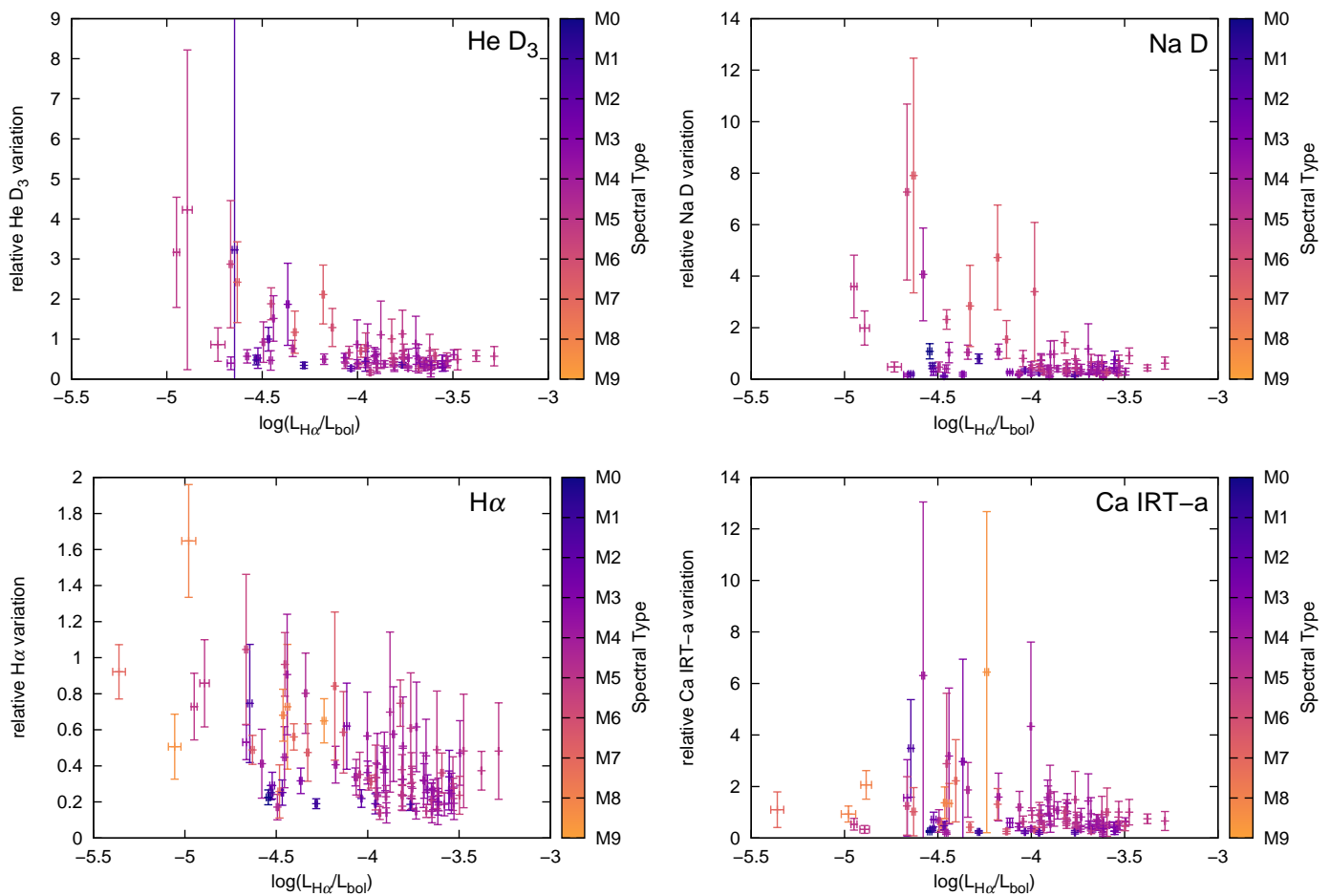}}
  \caption{Relative variation of chromospheric line $pEW'$ in H$\alpha$ active stars as a function of the normalised H$\alpha$ luminosity. Colours denote spectral types. Two outliers in He~D$_3$, one outlier in H$\alpha$, and two outliers in Ca~IRT-a are not shown for the sake of clarity.}
  \label{fig:pEW_var}
\end{figure*}

For better comparison across the different spectral types and activity levels, we divided the absolute variations by the average values as derived from the co-added spectra. The relative variations obtained in this way for H$\alpha$ active stars are shown as a function of the normalised H$\alpha$ luminosity in Fig.~\ref{fig:pEW_var}.

We find that the relative He~D$_3$ variation decreases with higher normalised H$\alpha$ luminosity with a Spearman correlation coefficient of $r_S = -0.49$. Two stars with $\log(L_{\mathrm{H}\alpha}/L_\mathrm{bol}) < -4.0$ are not shown in Fig.~\ref{fig:pEW_var} for reasons of clarity because their $pEW'_\mathrm{He~D_3} \approx 0\,${\AA} leads to a relative He~D$_3$ variation greater than $10$. The relative Na~D variation shows no clear correlation ($r_S = -0.24$). However, as eight of nine stars with a relative Na~D variation greater than $2$ have a normalised H$\alpha$ luminosity $\log(L_{\mathrm{H}\alpha}/L_\mathrm{bol}) < -4.0$, larger relative Na~D variations are more common at lower activity levels.

For the relative H$\alpha$ variation, we ignored one outlier caused by a poor spectrum of a star with only seven observations. With a moderate correlation coefficient of $r_S = -0.45$, we find that more active stars show a weaker relative H$\alpha$ variation, as was previously reported by \citet{2012PASP..124...14B} using the standard deviation of the $pEW$ as a measure of the variation. The anti-correlation between the relative Ca~IRT\nobreakdash-a variation and the normalised H$\alpha$ luminosity is weak ($r_S = -0.31$), but all stars with a relative Ca~IRT\nobreakdash-a variation greater than $2$ have a moderate normalised H$\alpha$ luminosity $\log(L_{\mathrm{H}\alpha}/L_\mathrm{bol}) < -4.0$. Two stars with $\log(L_{\mathrm{H}\alpha}/L_\mathrm{bol}) < -4.7$ have a $pEW'_\mathrm{Ca~IRT-a} \approx 0\,${\AA} leading to a relative Ca~IRT\nobreakdash-a variation greater than $10$, and are therefore not shown in Fig.~\ref{fig:pEW_var}.

On the whole, we find that the relative variations of He~D$_3$ and H$\alpha$ are moderately anti-correlated with the normalised H$\alpha$ luminosity, whereas there is only a weak anticorrelation for the relative variations of Na~D and Ca~IRT\nobreakdash-a. This anti-correlation is plausible because more active stars likely have a larger number of active regions, and consequently, the effect of a single active region appearing or disappearing on the disk-integrated spectrum is lower. The weaker anti-correlation in the cases of Na~D and Ca~IRT\nobreakdash-a may be a consequence of biases caused by the choice of the reference stars, as discussed in Sect.~\ref{subsubsection.sptdpnd.chromospheric}. The Na~D variation may also be affected by telluric emission in the line window in some spectra.

We do not show the relative variations of He~10833 and Pa$\beta$ because they do not show any correlation with the normalised H$\alpha$ luminosity, although the spectra with telluric contamination in the He~10833 and Pa$\beta$ line windows were not used to calculate the variations. The relative variations of the photospheric absorption band indices are not correlated with the normalised H$\alpha$ luminosity and therefore not shown either.

\subsection{Periodicities of activity indicators}
\label{subsection.periodicities}
In this section, we calculate periodicities of the activity indicators for the 331 stars in our sample and compare these candidate periods with known rotation periods. To remove outliers caused by flaring events or poor spectra, we rejected chromospheric line $pEW'$ and photospheric absorption band index values that differed from the respective average value by more than twice the standard deviation ($2\sigma$ clipping). For the He~10833 and Pa$\beta$ lines, we also rejected spectra with telluric contamination of the line window, as described in Sect.~\ref{subsection.pEW}. We then calculated the generalised Lomb-Scargle (GLS) periodogram \citep{2009A&A...496..577Z} for each activity indicator. The best-fit frequency for each indicator is the frequency with the maximum power in the GLS periodogram. We iteratively calculated the second- and third-best-fit frequencies from the GLS periodogram of the residuals of the best-fit frequency sine curve. Since most stars were observed at most once per night, we limited our calculations to frequencies below $1\,\mathrm{d}^{-1}$. The best-fit, second-best-fit, and third-best-fit frequencies for each star and each indicator are given in Table~\ref{table:results}.

Long-term activity cycles in M dwarfs have been found in the range 6$-$7\,yr \citep{2012A&A...541A...9G,2013ApJ...764....3R,2016A&A...595A..12S} and therefore likely cannot be seen in our data set, which covers only $2.4\,$yr. However, as the star rotates, different active regions become visible over time. The activity indicators may therefore vary with the rotation period $P_\mathrm{rot}$, which is known for 154 stars in our sample \citep{2007AcA....57..149K,2012AcA....62...67K,2012PASP..124..545H,2015ApJ...801..106H,2015MNRAS.452.2745S,2015ApJ...812....3W,2016ApJ...821...93N,2016A&A...595A..12S,2018arXiv181003338D}. Of these, 133 stars have a $P_\mathrm{rot} > 1\,$d, to which we can compare our calculated frequencies. We find 66 stars for which one of the three best-fit frequencies corresponds to $P_\mathrm{rot}$ within three times their respective errors in at least one indicator. Some of these matches may be by chance because only for 15 of these stars $P_\mathrm{rot}$ appears as one of the three best-fit frequencies in at least three indicators. As some active regions might disappear and new active regions might appear at different locations during the span of our observations, we cannot rule out that more stars show a semiperiodic variability of activity indicators induced by rotation over shorter time spans. However, as high-cadence observations have been conducted only for a small subset of the CARMENES sample, a period search using a shorter time baseline is beyond the scope of this paper.

\begin{figure}
  \resizebox{\hsize}{!}{\includegraphics{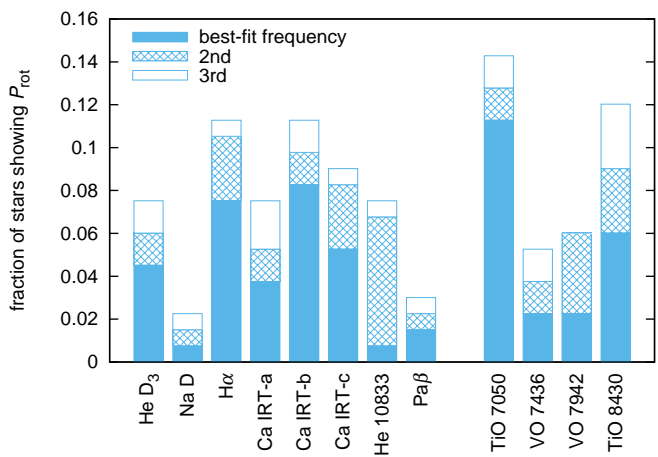}}
  \caption{Fraction of the 133 stars with $P_\mathrm{rot} > 1\,$d for which the best-fit, second-best-fist, or third-best-fit frequency in the chromospheric line $pEW'$ or photospheric absorption band index measurements corresponds to the rotation period $P_\mathrm{rot}$.}
  \label{fig:Prot_matches}
\end{figure}

For each indicator, we counted how many stars have a best-fit, second-best-fit, or third-best-fit frequency that corresponds to $P_\mathrm{rot}$ within three times their respective errors. A histogram is shown in Fig.~\ref{fig:Prot_matches}. We find that the TiO~7050 index varies with the rotation period most commonly, with 19 stars for which one of our three best-fit frequencies corresponds to $P_\mathrm{rot}$. $pEW'_{\mathrm{H}\alpha}$, $pEW'_{\mathrm{Ca~IRT-b}}$, and the TiO~8430 index are the only other indicators that show $P_\mathrm{rot}$ as one of the three best-fit frequencies for more than 10\% of the 133 stars with $P_\mathrm{rot} > 1\,$d.

%-------------------------------------------------------------------------------------------------------------------------------------------
\section{Summary}
\label{section.summary}
CARMENES searches for radial velocity variations of M dwarfs in the order of $1\,\mathrm{m\,s}^{-1}$ induced by planets orbiting in their liquid-water habitable zones. In this work, we have measured pseudo-equivalent widths of the He~D$_3$, Na~D, H$\alpha$, Ca~IRT, He~10833, and Pa$\beta$ lines for 331 M dwarfs using a spectral subtraction technique. Additionally, we measured indices of two photospheric TiO and two photospheric VO absorption bands. Our results improve our understanding of stellar activity in the CARMENES sample of M dwarfs.

We used the stars with the longest known rotation period as reference stars for each spectral subtype and found no stars with considerably stronger H$\alpha$ absorption than these reference stars. This implies that all our sample stars have at least a minimum level of chromospheric inactivity because we would expect to measure stronger H$\alpha$ excess absorption at low activity levels if the reference star had a completely inactive chromosphere \citep{1979ApJ...234..579C}. We defined stars with $pEW'_{\mathrm{H}\alpha} < -0.3\,${\AA} as H$\alpha$ active and found that 29\% of our sample stars are H$\alpha$ active. The fraction of H$\alpha$ active stars is below 20\% for spectral subtype M3.5\,V and earlier and significantly increases at later spectral subtypes. In addition, the normalised H$\alpha$ luminosity of typical H$\alpha$ active stars increases with spectral subtype up to M5.0\,V. This confirms the results of previous M dwarf activity studies by \citet{2012AJ....143...93R}, \citet{2015ApJ...812....3W}, and \citet{2018A&A...614A..76J}.

We showed that the He~D$_3$ line is strongly correlated with H$\alpha$, as was previously reported by \citet{2002AJ....123.3356G}. The Na~D doublet and the Ca~IRT show a weaker correlation because their photospheric components are too different for stars of the same spectral subtype and are therefore not completely removed by our spectral subtraction technique. He~10833 and Pa$\beta$ are challenging to measure because of telluric contamination and because they are not clearly correlated with the other indicators. The unclear correlation suggests that the He~10833 line is not very sensitive to activity, and this increases its usefulness as a tracer of exoplanet atmospheres \citep[e.g.][]{2018Natur.557...68S,2018Sci...362.1388N,2018A&A...620A..97S}.

We found that the photospheric TiO~7050 and TiO~8430 indices of H$\alpha$ active stars have lower values than the indices of H$\alpha$ inactive stars of the same spectral subtype. The VO~7436 and VO~7942 indices do not show this separation. We showed that this behaviour can be explained by rotational and Zeeman broadening, which are stronger in active stars.

We quantified the variation of the activity indicators for all stars in the sample. In units of the respective $pEW'$ as derived from co-added spectra, the variations in the H$\alpha$ and He~D$_3$ lines among multiple observations are smaller for more active stars. The anti-correlation between the normalised H$\alpha$ luminosity and the relative variations in Na~D and Ca~IRT\nobreakdash-a lines is weaker, whereas the variations in He~10833 and Pa$\beta$ show no correlation.

We calculated periodicities in the activity indicators of each star using GLS periodograms and compared the three best-fit frequencies with known rotation periods $P_\mathrm{rot}$. The TiO~7050 index, $pEW'_{\mathrm{H}\alpha}$, $pEW'_{\mathrm{Ca~IRT-b}}$, and the TiO~8430 are most likely to vary with the rotation period. Of the 133 stars with $P_\mathrm{rot} > 1\,$d, 66 show the rotation period in at least one activity indicator. Because the other 67 stars could show the rotation period in the activity indicators during parts of the time baseline, a semi-periodic activity variation of a different origin might not be present all the time either. However, this semi-periodic variation could still lead to periodic changes in the measured radial velocity, which might be misinterpreted as induced by a planet. Only 15 stars show the rotation period in at least three activity indicators. A careful study of individual stars is necessary to understand why the rotation period may appear in some but not all indicators, but this is beyond the scope of this paper, which is focused on the full CARMENES sample.

%-------------------------------------------------------------------------------------------------------------------------------------------
\begin{acknowledgements}
CARMENES is an instrument for the Centro Astron\'omico Hispano-Alem\'an de Calar Alto (CAHA, Almer\'{\i}a, Spain). CARMENES is funded by the German Max-Planck-Gesellschaft (MPG), the Spanish Consejo Superior de Investigaciones Cient\'{\i}ficas (CSIC), the European Union through FEDER/ERF FICTS-2011-02 funds, and the members of the CARMENES Consortium (Max-Planck-Institut f\"ur Astronomie, Instituto de Astrof\'{\i}sica de Andaluc\'{\i}a, Landessternwarte K\"onigstuhl, Institut de Ci\`encies de l'Espai, Institut f\"ur Astrophysik G\"ottingen, Universidad Complutense de Madrid, Th\"uringer Landessternwarte Tautenburg, Instituto de Astrof\'{\i}sica de Canarias, Hamburger Sternwarte, Centro de Astrobiolog\'{\i}a and Centro Astron\'omico Hispano-Alem\'an),  with additional contributions by the Spanish Ministry of Science, the German Science Foundation through the Major Research Instrumentation Programme and DFG Research Unit FOR2544 ``Blue Planets around Red Stars'', the Klaus Tschira Stiftung, the states of Baden-W\"urttemberg and Niedersachsen, and by the Junta de Andaluc\'{\i}a.
\end{acknowledgements}

\bibliographystyle{aa}
\bibliography{sample_activity}

\begin{appendix}
 \section{Table of results}
 \longtab[1]{
\scriptsize
\begin{landscape}
\begin{longtable}{l l l r c l l c c l c c c c c c}
\label{table:results}\\
\caption{Identification, common name, Gliese number, spectral type (SpT), stellar population (Pop.), rotational velocity, rotation period, normalised H$\alpha$ luminosity, $pEW'_{\mathrm{H}\alpha}$, absolute H$\alpha$ variation, and the three best-fit frequencies for H$\alpha$.}\\
   \hline
   \hline
   \noalign{\smallskip}
No. & ID & Name & Gl/GJ & SpT & Ref. & Pop. & $v\sin i$ & $P_\mathrm{rot}$ & Ref. & $\log(L_{\mathrm{H}\alpha}/L_\mathrm{bol})$ & $pEW'_{\mathrm{H}\alpha}$ & abs. H$\alpha$ variation & H$\alpha$ best-fit freq. & 2nd & 3rd \\
    &    &      &       &     &     &     & [km\,s$^{-1}$] & [d]         &     &                                             & [\AA]                     & [\AA]                    & [d$^{-1}$]               & [d$^{-1}$] & [d$^{-1}$]\\
\noalign{\smallskip}
    \hline
    \noalign{\smallskip}		
 \endfirsthead
\caption{Identification, common name, Gliese number, spectral type (SpT), stellar population (Pop.), rotational velocity, rotation period, normalised H$\alpha$ luminosity, $pEW'_{\mathrm{H}\alpha}$, absolute H$\alpha$ variation, and the three best-fit frequencies for H$\alpha$.}\\ 
  \hline
  \hline
  \noalign{\smallskip}		
No. & ID & Name & Gl/GJ & SpT & Ref. & Pop. & $v\sin i$ & $P_\mathrm{rot}$ & Ref. & $\log(L_{\mathrm{H}\alpha}/L_\mathrm{bol})$ & $pEW'_{\mathrm{H}\alpha}$ & abs. H$\alpha$ variation & H$\alpha$ best-fit freq. & 2nd & 3rd \\
    &    &      &       &     &     &     & [km\,s$^{-1}$] & [d]         &     &                                             & [\AA]                     & [\AA]                    & [d$^{-1}$]               & [d$^{-1}$] & [d$^{-1}$]\\
 \noalign{\smallskip}
  \hline
  \noalign{\smallskip}
  \endhead
  \noalign{\smallskip}
  \hline
  \endfoot
1 & J00051+457 & GJ 2 & 2 & M1.0\,V & PMSU & YD & \ldots & 15.5$\pm$0.7 & DA18 & \ldots & --0.064$\pm$0.008 & 0.048$\pm$0.007 & 0.0556$\pm$0.0002 & 0.1283$\pm$0.0003 & 0.7301$\pm$0.0003 \\
2 & J00067--075 & GJ 1002 & 1002 & M5.5\,V & PMSU & D & \ldots & \ldots & \ldots  & \ldots & 0 & 0.13$\pm$0.03 & 0.0459$\pm$0.0003 & 0.9866$\pm$0.0003 & 0.8991$\pm$0.0003 \\
3 & J00162+198E & LP 404--062 & 1006B & M4.0\,V & AF15 & TD-D & \ldots & 105$\pm$44 & DA18 & \ldots & 0.052$\pm$0.011 & 0.14$\pm$0.08 & 0.2681$\pm$0.0007 & 0.5012$\pm$0.0004 & 0.1764$\pm$0.0013 \\
4 & J00183+440 & GX And & 15A & M1.0\,V & AF15 & D & \ldots & \ldots & \ldots  & \ldots & --0.073$\pm$0.007 & 0.031$\pm$0.004 & 0.0070$\pm$0.0002 & 0.2717$\pm$0.0003 & 0.5607$\pm$0.0004 \\
5 & J00184+440 & GQ And & 15B & M3.5\,V & PMSU & YD & \ldots & \ldots & \ldots & \ldots & --0.033$\pm$0.012 & 0.048$\pm$0.005 & 0.0024$\pm$0.0002 & 0.3850$\pm$0.0002 & 0.9727$\pm$0.0003 \\
6 & J00286--066 & GJ 1012 & 1012 & M4.0\,V & PMSU & D & \ldots & \ldots & \ldots & \ldots & 0.122$\pm$0.016 & 0.06$\pm$0.03 & 0.2750$\pm$0.0003 & 0.6070$\pm$0.0004 & 0.2637$\pm$0.0006 \\
7 & J00389+306 & Wolf 1056 & 26 & M2.5\,V & AF15 & D & \ldots & 50.2$\pm$1.3 & DA18 & \ldots & 0.036$\pm$0.007 & 0.044$\pm$0.017 & 0.8385$\pm$0.0007 & 0.2051$\pm$0.0004 & 0.8922$\pm$0.0008 \\
8 & J00570+450 & G 172--030 & \ldots & M3.0\,V & Lep13 & YD & \ldots & \ldots & \ldots & \ldots & --0.081$\pm$0.011 & 0.13$\pm$0.05 & 0.0380$\pm$0.0005 & 0.0272$\pm$0.0007 & 0.1121$\pm$0.0004 \\
9 & J01013+613 & GJ 47 & 47 & M2.0\,V & PMSU & D & \ldots & \ldots & \ldots & \ldots & --0.069$\pm$0.008 & 0.15$\pm$0.11 & 0.6223$\pm$0.0001 & 0.9434$\pm$0.0001 & 0.0538$\pm$0.0001 \\
10 & J01019+541 & G 218--020 & 3069 & M5.0\,V & PMSU & D & 30.6$\pm$3.1 & 0.278$\pm$0.001 & DA18 & --3.764$\pm$0.004 & --5.74$\pm$0.06 & 3.5$\pm$1.8 & 0.1302$\pm$0.0003 & 0.8110$\pm$0.0002 & 0.5330$\pm$0.0002 \\
\end{longtable}
\tablefoot{The full table with all activity indicators is available at the CDS. Stellar populations are adopted from \citet{CC16} and abbreviated as follows: YD: young disc, D: thin disc, TD-D: transition between thin
and thick disc, TD: thick disc. $v\sin i$ are adopted from from \citet{2018A&A...612A..49R} if not indicated otherwise.}
\tablebib{PMSU: Palomar/Michigan State University survey \citep{1995AJ....110.1838R}; Lep13: \citet{2013AJ....145..102L}; AF15: \citet{2015A&A...577A.128A}; DA18: \citet{2018arXiv181003338D}.}
\end{landscape}
}
\end{appendix}
\end{document}